\documentclass[onecolumn,superscriptaddress,aps,preprintnumbers,amsmath,amssymb]{revtex4-1}
\UseRawInputEncoding
\usepackage{graphicx}
\usepackage{dcolumn}
\usepackage{bm}
\usepackage{algorithm2e}
\usepackage{xfrac}
\usepackage{braket}
\usepackage{mathtools,amssymb,lipsum} 
\usepackage[utf8]{inputenc}
\renewcommand{\figurename}{{\bf{Fig.}}}
\renewcommand\tablename{{\bf{Table}}}
\usepackage{color,soul}
\begin{document} 
\author{Suvadip Masanta}
\affiliation{Department of Physical Sciences, Bose Institute, EN 80, Sector V, Bidhannagar, Kolkata 700091, India}
\author{Chumki Nayak}
\affiliation{Department of Physical Sciences, Bose Institute, EN 80, Sector V, Bidhannagar, Kolkata 700091, India}
\author{Premananda Chatterjee}
\affiliation{Department of Condensed Matter and Materials Physics, S. N. Bose National Centre for Basic Sciences, Kolkata 700106, India}
\author{Atindra Nath Pal}
\affiliation{Department of Condensed Matter and Materials Physics, S. N. Bose National Centre for Basic Sciences, Kolkata 700106, India}
\author{Indrani Bose}
\affiliation{Department of Physical Sciences, Bose Institute, EN 80, Sector V, Bidhannagar, Kolkata 700091, India}
\author{Achintya Singha}
\email{achintya@jcbose.ac.in}
\affiliation{Department of Physical Sciences, Bose Institute, EN 80, Sector V, Bidhannagar, Kolkata 700091, India}

\begin{abstract}
Heterobilayers formed by stacking two-dimensional atomic crystals are particularly promising for low-dimensional semiconductor optics, as they host interlayer excitons—bound states of electrons and holes residing in different layers. They inherit the valley-contrasting physics of the individual monolayers, leading to a range of unique properties that distinguish them from other solid-state nanostructures. Here, we propose a novel route for the generation of interlayer excitons based on the synthesis of a transition metal dichalcogenide bilayer alloy material, WS$_{2x}$Se$_{2(1-x)}$. Using piezoelectric force microscopy, we demonstrate the existence of an internal electric field oriented in the out-of-plane direction. Interlayer excitons have so far been mostly observed in heterostructures with a type-II band alignment. In the presence of an internal electric field, a similar alignment occurs in the alloy bilayer resulting in an efficient generation of interlayer excitons. Photoluminescence spectroscopy measurements involving circularly polarised light come up with key observations like a negative degree of circular polarization of the interlayer excitons which increases as a function of temperature. A simple theoretical model provides a physical understanding of the major experimentally observed features. With experimentally fitted parameter values, the dominant contribution to the degree of circular polarization is shown to arise from spin polarization and not from valley polarization, a consequence of the spin-valley-layer coupling characteristic of a TMDC bilayer. The room-temperature interlayer excitonic transition in bilayer TMDCs has key implications for fundamental physics, including Bose-Einstein condensation and high-temperature superfluidity, while enabling advanced valleytronic and quantum information functionalities.
\end{abstract}
\title{Alloying as a new route to generating interlayer excitons}\maketitle
\subsection*{Introduction}
 Transition metal dichalcogenides (chemical formula unit MX$_2$, M = Mo, W, X = S, Se) are a class of semiconducting materials exhibiting a number of fascinating physical properties in their monolayer and few-layer forms \cite{Schaibley2016, Xu2014, Rivera2018}. The energy bands of the TMDC materials exhibit two prominent valleys at the $K$ and $-K$ points of the Brillouin zone, with the respective energy levels related via time reversal symmetry, $E(\textbf{k}, \uparrow)=E(-\textbf{k}, \downarrow)$. In contrast to the TMDC monolayer which is inversion asymmetric and a direct band gap semiconductor, the pristine bilayer is an indirect band gap semiconductor with global inversion symmetry (IS) in its 2H stacking configuration. The stacking configuration is marked by a 180$^{\circ}$ relative rotation between the upper and lower layers which interchanges the $K$ and $-K$ valleys between the layers keeping the spin configurations unchanged \cite{Rivera2018}. In a direct $K$-$K$ transition, an electron is optically excited from the valence band (VB) to the conduction band (CB) in the $K$ valley. The electron in the CB and the positively charged hole in the VB form a bound state characterised as an exciton. The efficient generation of an interlayer exciton (IX) in a few-layer material is facilitated by a type-II band alignment in which the energy degeneracy between the layers is removed due to the breaking of global IS. Such band alignments occur naturally in heterostructures which combine two different TMDC materials. In a type-II band alignment, the VB maximum and the CB minimum belong to different layers (Fig. 1(b)). On the photogeneration of an A exciton in either monolayer, an ultrafast interlayer charge transfer takes place on a femtosecond  timescale resulting in the electron (hole) relaxing to the CB (VB) minimum (maximum). The subsequent binding of the electron and the hole results in the formation of an IX.  
The excitons, both intralayer and interlayer, have finite lifetimes due to the recombination of the electron-hole pair. The process is marked by the emission of light constituting a photoluminescence (PL) signal. The recombination time and hence the lifetime of an exciton depends on the overlap of the electron and hole wavefunctions. This overlap is much reduced in the case of an IX due to the spatial separation of the electron and hole in two different layers. Accordingly, the IX has a lifetime ($\sim$ ns) which is much larger than the lifetime ($\sim$ ps) of the intralayer exciton (IL), enhancing its application potential in exciton-based devices \cite{Rivera2018, Jiang2021, doi:10.1126/science.aaw4194}. Another interesting property of the IX is its possession of a permanent electric dipole moment which can be tuned via electrical gating \cite{Rivera2018}. Furthermore, the existence of repulsive dipole-dipole interactions allows for the exploration of fascinating quantum many body effects like Bose-Einstein condensation \cite{Rivera2018}. As recent reports suggest, optical studies of interlayer excitons have mostly focused on TMDC heterostructures, the fabrication methods of which are quite challenging \cite{Kunstmann2018, Chatterjee2023, PhysRevB.105.L241406, Blundo2024}. Additionally, the quality of the formed interfaces significantly influences the coupling of the layers, posing difficulties in controlling the excitonic properties. In this scenario, naturally stacked bilayers can be a better alternative as they are devoid of contaminants between the layers, thus yielding samples of exceptionally high quality with enhanced PL emission efficiency.
\begin{figure*}[t!]
\centering
\includegraphics[width=1\linewidth]{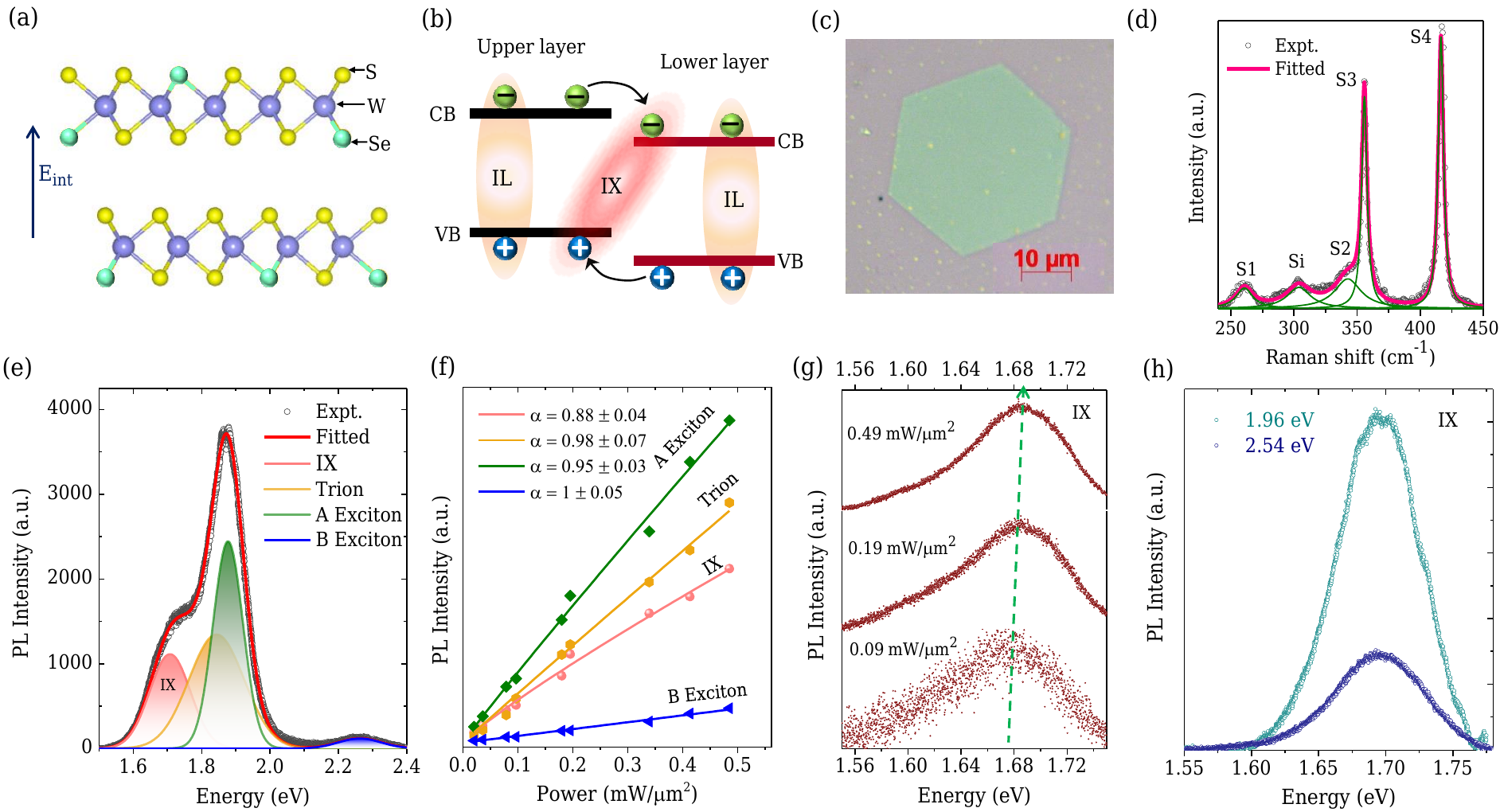}
\caption{Schematic illustration of bilayer alloy WSSe and primary characterizations. (a) Side view of bilayer WSSe structure,(b) sketch of type-II band alignment of WSSe bilayer, (c) optical image, (d) Raman spectrum, (e) PL spectrum at RT, (f) optical power dependent PL intensity plot (symbols) for all transitions. Solid lines represent fits to the data using a power-law relationship (intensity $\propto$ Power$^{\alpha}$), (g) power-dependent PL spectra of the interlayer exciton (IX), showing a blue shift of $\sim$ 10 meV with increasing excitation power, as indicated by the dashed arrow. This behavior is a characteristic signature of interlayer excitons. (h) excitation energy dependent (2.54 eV and 1.96 eV) emissions of IX.
}
\end{figure*}
\par
In this study, we demonstrate a new route, namely alloying, to the generation of interlayer excitons in a tungsten-based bilayer material. We synthesize the TMDC bilayer alloy material WS$_{2x}$Se$_{2(1-x)}$, ($x=0.88$) and establish, using piezoelectric force microscopy (PFM), the existence of an internal electric field oriented in the out-of-plane direction. The PFM study is the first of its kind carried out on a tungsten-based TMDC material. The PL emission spectrum gives clear evidence of the intralayer A exciton and a lower energy IX. The identity of the latter is confirmed by laser power-dependent, polarization-dependent, excitation energy-dependent, and gate voltage-dependent PL responses. The interesting observations related to the IX are: the circularly polarized emission helicity is opposite to that of the optical excitation generating the A exciton, the PL emission energy is blue-shifted as a function of temperature, which is a signature of repulsive dipole–dipole interactions between interlayer excitons, and the modulus of the degree of circular polarization increases with temperature. The last feature, not reported before, is of significant interest in the context of room temperature (RT) device applications. 
A physical understanding of the experimentally-observed features is provided by a theoretical model incorporating the spin-valley-layer coupling and the interlayer hopping of the VB hole \cite{Gong2013, Jones2014}. In particular, we show, using experimentally fitted parameter values, that the dominant contribution to the degree of circular polarization comes from spin polarization and not from valley polarization. 
\subsection*{Results}
\subsubsection{Material synthesis and primary characterizations}
The WSSe alloy was synthesized on a SiO$_2$/Si substrate using the chemical vapor deposition (CVD) technique as described in Section A, Supplemental Material (SM). Figure 1(a) schematically depicts the side view of alloy WSSe structure. The optical microscope image of the as-grown sample, presented in Fig. 1(c), shows a hexagonal-shaped WSSe flake on the substrate. Elemental analysis using energy-dispersive X-ray (EDX) spectroscopy was conducted to determine the atomic composition of W, S, and Se. The estimated percentages of W, S, and Se from the EDX spectrum, presented in Fig. S2(a) of the SM , are 33.3\%, 58.3\%, and 8.4\%, respectively. Topographical analysis using atomic force microscopy, shown in Fig. S2(b) and (c), reveal a flake thickness of 1.5 nm, confirming the bilayer structure. The Raman spectrum in Fig. 1(d) exhibits prominent peaks at 260.7 cm$^{-1}$ (S1), 342.6 cm$^{-1}$ (S2), 356 cm$^{-1}$ (S3), and 416 cm$^{-1}$ (S4). Based on the previous studies, these peaks (S1, S3, S4) are assigned to the A$_{1g}$ (Se-W), E$_{2g}$ (S-W), and A$_{1g}$ (S-W) vibrational modes, respectively \cite{doi:10.1021/acs.nanolett.5b03662, Zribi_2023}. The S2 mode may arise from alloying effects or be analogous to the 2LA mode of WS$_2$ \cite{PhysRevB.100.235438, Berkdemir2013}.
\par
The RT PL spectrum of the bilayer flake under 2.54 eV excitation is shown in Fig. 1(e). A temperature-dependent PL study of bilayer WSSe, ranging from 300 K to 90 K, is discussed in a later section. Figure S4 presents PL spectra from sixteen different positions across two additional samples from separate batches, demonstrating the homogeneity and reproducibility of our CVD-grown material. Though the alloy bilayer material has an indirect gap, the PL spectrum has dominant contributions from the direct optical transitions at the $K$/$-K$ valleys, as indirect optical transitions depend on phonon/defect scatterings to conserve momentum. 
The PL emission features correspond to the recombination of the intralayer A and B excitons and also the A trion, with energy peaks located at 1.877 eV, 2.263 eV, and 1.843 eV respectively. The A (B) exciton is associated with the transition from the upper (lower) spin-split VB. Notably, a prominent emission peak (denoted as IX) is observed at a lower energy of 1.710 eV. This feature has not been previously observed in pristine WS$_2$, WSe$_2$ monolayers \cite{doi:10.1021/acs.nanolett.5b03662, Ernandes2021}. We attribute this peak to the formation and subsequent recombination of an interlayer (IX) exciton. Before we discuss the experimental evidence for this, we discuss briefly the nature of the other prominent peaks appearing in the PL spectrum. In the case of our alloy bilayer sample, the difference in the peak positions of the A and B excitons is $\sim$ 0.386 eV, which is close to the almost constant value  0.4 eV of the A-B exciton splitting, as reported in  experimental PL studies on multilayer WX$_2$, (X=S, Se) with the number of layers running from 1 to 4 \cite{Zeng2013, doi:10.1021/nn305275h}. In the case of monolayer WX$_2$, the magnitude of the exciton spin-splitting is determined by the VB spin-valley coupling strength $\lambda_v$ (2$\lambda_v=0.4$ eV) \cite{Zeng2013}. The near-constancy of the magnitude of the exciton splitting, irrespective of the number of layers in the sample, indicates the suppression of interlayer hopping in the tungsten-based TMDC materials. This is also true for our alloy bilayer sample.
\par
The trion is a charged exciton, described as a bound state of the neutral A exciton and an electron (A$^-$) or a hole (A$^+$). The trion peak has an energy lower than that of the neutral exciton, within a range of  values, $\sim$ 0.03-0.45 eV from the peak energy of the A exciton \cite{10.1063/1.4983285, Jones2014, https://doi.org/10.1002/pssr.201510224}. In the case of our alloy bilayer, the energy gap of 0.034 eV between the A exciton and trion peaks falls within the reported range. The energy gap between the A and IX exciton peaks is 0.167 eV. The IX exciton is weakly-bound compared to the A exciton since the separation distance between the electron and  hole is larger in the case of the former. The peak in the PL emission spectrum thus occurs at a lower energy. Figure 1(f) presents the PL intensity (I) as a function of excitation power (P) derived from the power-dependent PL spectra at RT shown in Fig. S3. A power-law fit of the data (solid line) in the figure, using the relation $I \propto P^{\alpha}$, yields the exponent ($\alpha$) values as: 0.98 $\pm$ 0.07 (A trion), 0.95 $\pm$ 0.05 (A exciton) and 1.0 $\pm$ 0.05 (B exciton), which are consistent with the previously obtained estimates for TMDC materials \cite{doi:10.1021/nn5059908, 10.1063/1.4983285}. For the IX peak, the exponent $\alpha$ has the value $\alpha$ = 0.88 $\pm$ 0.05, i.e., nearly linear behaviour. Figure 1(g) displays the power-dependent PL of IX excitons at 90 K (PL spectra for various excitation power densities ranging from 0.02 mW/$\mu$m$^2$ to 0.49 mW/$\mu$m$^2$ are shown in Fig. S5). The dashed arrow indicates a blue shift of the PL peak as the excitation power increases. The IX exciton carries a static electric dipole moment pointing in the out-of-plane direction. With increasing excitation power, the IX exciton density increases resulting in a blue shift of the peak energy due to repulsive dipole-dipole interactions. From the lowest to the highest power used, the blue shift in the peak energy is around 10 meV. The repulsive interaction, combined with the enhanced lifetime of the IX excitons, allow for the exploration of many body phenomena like excitonic Bose-Einstein condensation and superconductivity \cite {Jiang2021, Rivera2015}. The interlayer nature of IX emission is further corroborated by excitation-energy dependent PL spectroscopy, utilizing two distinct excitation energies: 2.54 eV and 1.96 eV, under identical experimental conditions. A pronounced enhancement in IX PL emission at 100 K is observed when the excitation energy (1.96 eV) is resonant with the intralayer (A) exciton, as compared to the emission observed using 2.54 eV excitation (Fig. 1(h)). This enhancement indicates that the IX exciton is formed through a charge transfer (energy relaxation) after the generation of an intralayer exciton  in either monolayer \cite{Hsu2018, PhysRevB.105.L241406, Nagler_2017}. As mentioned in the introduction, a type-II band alignment is a prerequisite for the efficient generation of IX excitons. Through our PFM study, described in the next section, we provide a confirmatory evidence for the existence of an internal electric field with an out-of-plane direction in our bilayer sample. The field generates a potential difference between the two layers of the bilayer giving rise to a type-II band alignment. In Section 3, we show, through PL studies, that the IX excitons exhibit a negative degree of circular polarization, a feature which has been demonstrated earlier in the case of the interlayer excitonic emission in WSe$_2$/MoSe$_2$ heterobilayer \cite{Hsu2018}. Furthermore, applying a vertical gate bias across the bilayer sample demonstrates that the emission energy of the interlayer exciton is tunable \cite{Rivera2015}. The device fabrication for this measurement is described in SM. In SM, Fig. S6(b), we show the variation in peak energy of the IX in our bilayer sample by changing the gate voltage from –60 V to +60 V, (some representative PL spectra are shown in Fig. S6(a)). Each branch in Fig. S6(b) exhibits a linear shift in the peak emission energy of the interlayer exciton as a function of gate voltage. The orientation of the IX dipole is primarily controlled by the electric field induced by the applied gate voltage, leading to a redshift in the peak energy for both negative and positive gate voltages. In Section H of the SM, we include a brief description of how the A and IX exciton peaks respond to linearly polarized light using both 2.54 eV and 1.96 eV excitations. The degree of linear polarization is computed in each case, indicative of valley coherence \cite{Jones2014, doi:10.1073/pnas.1406960111}.

\subsubsection{PFM study}
In our as-synthesized alloy WSSe, we replaced a fraction of the chalcogen S atoms with Se on both sides of the central W atoms (as depicted schematically in Fig. 1(a)). This substitution introduces a charge imbalance in the vertical direction, due to the differing electronegativities of S and Se \cite{PhysRevB.109.115304}. This intrinsic charge imbalance induces a net electric dipole moment and hence an electric field in the vertical direction. In the presence of the field, a potential difference exists between the upper and lower layers giving rise to upward and downward shifts respectively of the electronic energy bands and resulting in a type-II band edge alignment (as shown in Fig. 1(b) and Fig. 3(d)). To experimentally verify the presence of this intrinsic out-of-plane electric field, we conducted PFM measurements in the vertical mode. The PFM is a powerful technique for the investigation of piezoelectric and ferroelectric phenomena at the nanoscale \cite{PhysRevB.109.115304, Lu2017, Damjanovic1998, https://doi.org/10.1002/adma.202204697}. The piezoresponse of the PFM measurements makes use of the inverse piezo effect involving the development of mechanical strain in response to an applied voltage. Ferroelectric materials are a subset of piezoelectric materials and exhibit spontaneous polarization due to the existence of permanent electric dipoles.
\begin{figure*}[t!]
 \centering
\includegraphics[width=0.6\linewidth]{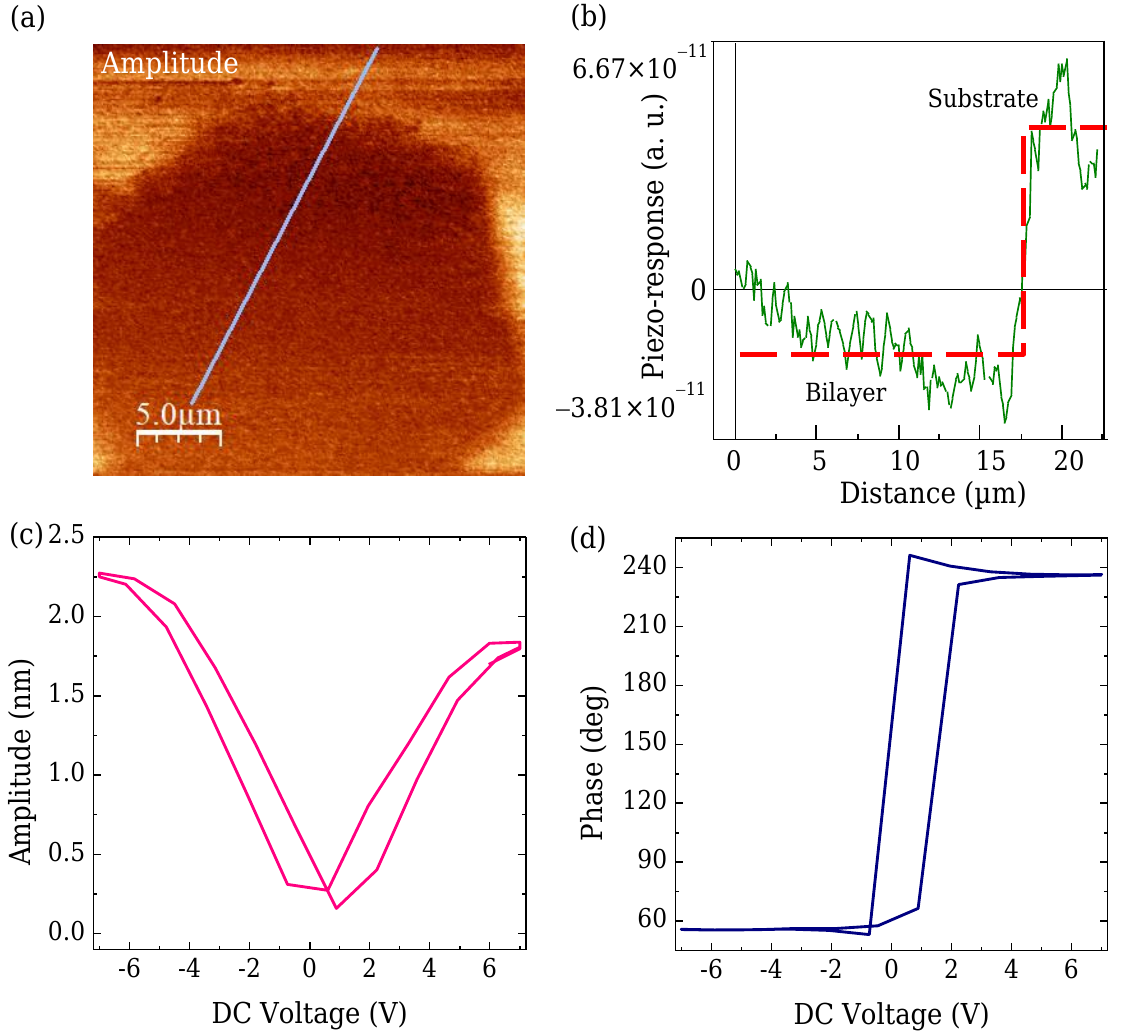}
\caption{ PFM measurement on bilayer WSSe. (a) Piezoelectric amplitude of hexagonal-shaped WSSe on SiO$_2$/Si substrate and (b) the piezoelectric line profile corresponding to the blue line on image (a). (c) The PFM amplitude and (d) phase of the bilayer WSSe flake as a function of DC bias voltage.}
\end{figure*}
\par
Figure 2(a) and S8(b) present the piezoelectric amplitude domain and phase image of the bilayer sample. The observed contrast between the sample and the SiO$_2$/Si substrate (in Fig. 2(a)), together with the piezoresponse profile (Fig. 2(b)) extracted from the marked line in Fig. 2(a), confirms that the piezoelectric response originates from the sample itself. The core principles underlying the PFM measurements are described in Section I of SM. 
The piezoresponse is described in terms of two quantities: the strain amplitude and the phase, with the phase providing information about the polarization direction. Figures 2(c) and (d) show the plots of the amplitude and phase respectively versus the applied DC voltage sweeping over the range of values -7 V to +7 V. An AC probe voltage rides on the sweeping DC voltage to keep simultaneous track of the sample characteristics. The butterfly shape of the amplitude curve is a characteristic feature of piezoelectric materials subjected to varying electric fields and reflects the strain response of the material. The phase plot shows a well-defined hysteresis loop with sharp polarization switches ($\approx$ 180$^\circ$) in a varying external field. The combination of the butterfly and phase hysteresis curves, marked by  local polarization switches, provides a direct evidence for the existence of an intrinsic out-of-plane electric dipole moment, and its associated electric field \cite{Lu2017, C1NR11099C, doi:10.1021/acsnano.8b02152, https://doi.org/10.1002/adma.202204697}.
\begin{figure*}[t!]
 \centering
\includegraphics[width=0.8\linewidth]{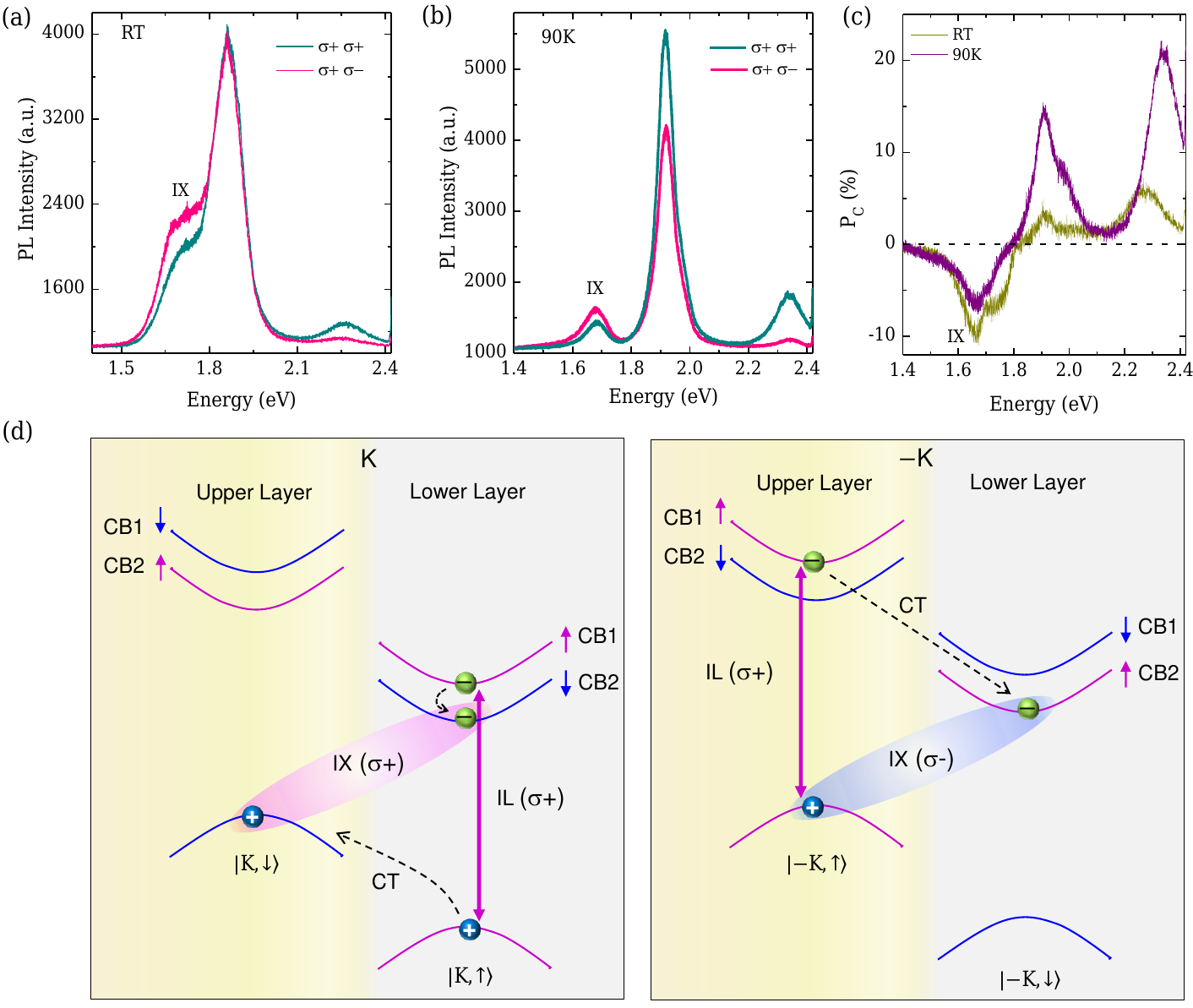}
\caption{Helicity-resolved PL of bilayer WSSe using $\sigma+$ excitation at 2.54 eV with $\sigma+$ and $\sigma-$ detection configurations. (a) Room temperature and (b) 90 K temperature. (c) The degree of circular polarization for (a) and (b). (d) Schematic representation of the energy diagram of bilayer WSSe and transitions.}
\end{figure*}
\subsubsection{PL study, degree of polarization and Model Hamiltonian}
Figures 3(a) and (b) present the helicity-resolved PL spectra obtained using $\sigma+$ (right circularly polarized) excitation from an off-resonance 2.54 eV laser at both RT and 90 K, respectively. To quantify the degree of circular polarization ($P_C$), we use the formula: $P_C= {I(\sigma+)-I(\sigma-)}/{I(\sigma+)+I(\sigma-)}$, where I($\sigma+$) and I($\sigma-$) represent the PL intensities measured under co- (right) and cross- (left) circularly polarized configurations, respectively \cite{Hsu2018, Mak2012}. At both the temperatures, the helicity of the IX is opposite to that of the A exciton. Figure 3(c) shows $P_C$ as a function of energy (eV). At RT, it is evident that both A and  B excitons exhibit a positive $P_C$ while interestingly, the IX transition demonstrates a significant negative $P_C$ (-10\%). A similar negative circular polarization is observed while exciting the sample with $\sigma-$ excitation, as shown in Fig. S10. To check the consistency of the results, we conduct helicity-resolved PL measurements on two additional samples and the results are shown in Fig. S11. This helicity-dependent optical response of the IX peak rules out its origin from indirect transitions which are known to be unpolarized \cite{doi:10.1126/science.aan8010, doi:10.1073/pnas.1406960111}. It is also observed that the magnitude of $P_C$ for the A and B excitons increases whereas that for the IX decreases, when the temperature is reduced from its RT value to 90 K (Fig. 3(c))]. We further conducted detailed temperature-dependent PL measurements on our bilayer sample (Fig. S12(a)). Figure 4(a) shows the color plot of the  PL spectra from RT down to 90 K. Figure 4(b) shows the variation of the PL intensity for the A exciton and IX as a function of temperature. The variation of the peak energy of the A exciton and IX versus temperature is shown in Fig. 4(c).
\begin{figure*}[t!]
 \centering
\includegraphics[width=0.7\linewidth]{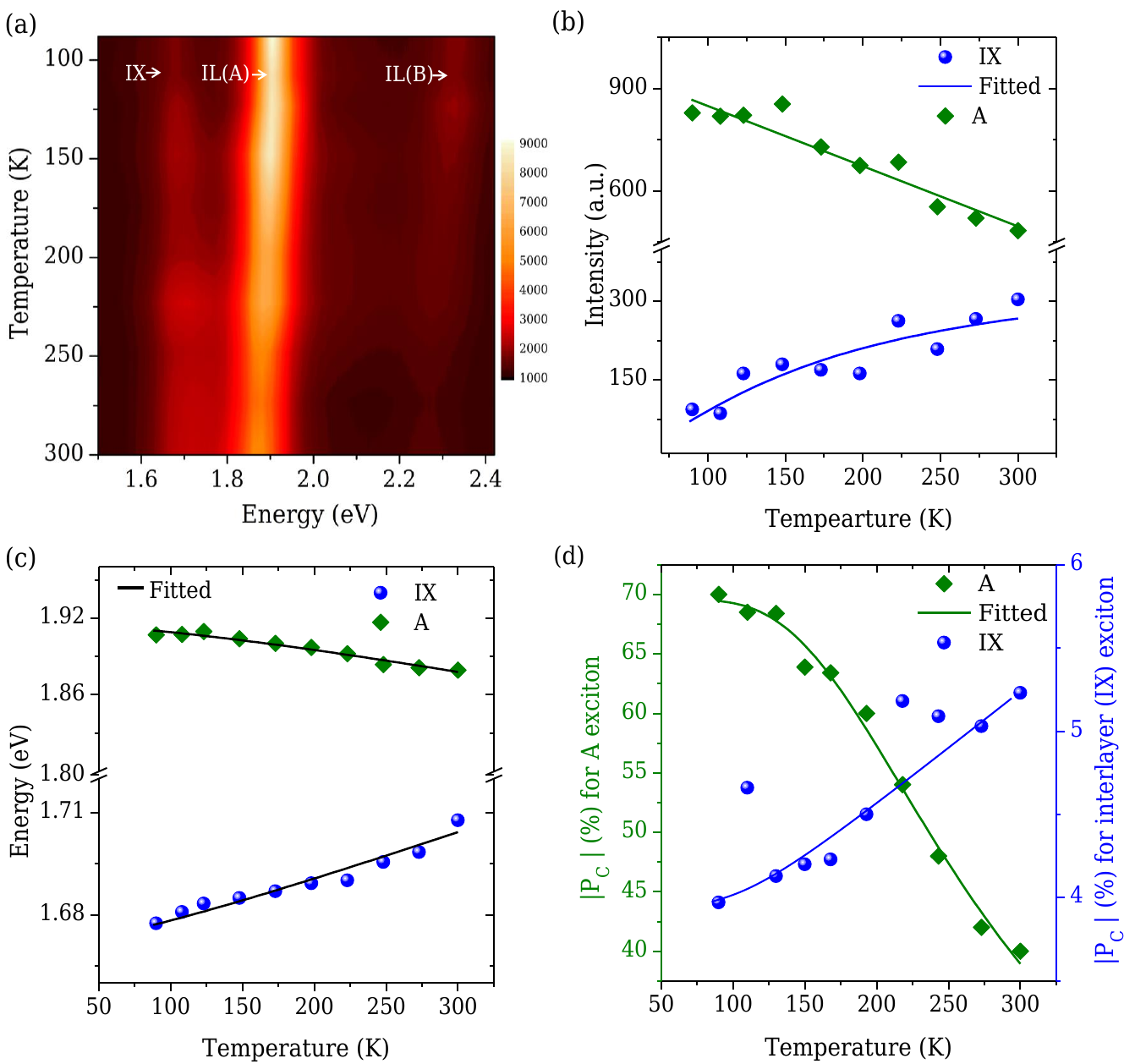}
\caption{Temperature-dependent PL study on bilayer WSSe. (a) Color plot of the PL intensity as function of energy and temperature. (b) Extracted PL intensity and (c) PL energy as a function of temperature for IX and intralayer A transitions. The evolution of the peak energies with temperature are fitted using the Varshni equation (black solid line). These experimental results shown in (a), (b), and (c) were performed utilizing 2.54 eV excitation. (d) Temperature dependence of the modulus of $P_C$(\%) for the A exciton and IX. These results are obtained from the helicity resolved PL study, using near on resonance excitation of 1.96 eV laser with $\sigma+$ helicity from the temperature range 300 K to 90 K, as depicted in Fig. S13(c).
}
\end{figure*}
\par
To explore the impact of excitation energy on circularly polarized emission, we performed helicity-resolved PL measurements using near on-resonance (1.96 eV) laser excitation (matching the A exciton energy) with $\sigma+$ helicity, over a temperature range of 90 K – 300 K (see Fig S13(a), (b)).
The calculated $P_C$ as a function of emission energy (eV) at various temperature is presented in Fig. S13(c). Similar to the earlier observation with off-resonance excitation (Fig. 3(c)), these results also indicate that the helicities of the A exciton and IX are opposite. Figure 4(d) depicts the $|P_{C}|$ (\%) for the A exciton and IX, as a function of temperature. For the A exciton, $|P_{C}|$ increased as the temperature decreased, reaching approximately 72\% at 90 K. In contrast, the interlayer exciton exhibited a decreasing $|P_{C}|$ as temperature decreased, consistent with Fig. 3(c). The observations suggest that temperature can serve as a tunable parameter for controlling the circular polarization of both the A exciton and IX. Furthermore, power-dependent helicity-resolved PL measurements under near on resonance excitation (with A exciton) at RT and 90 K (Fig. S14) reveal a potential avenue to enhance the polarization degree for both the exciton types. A robust polarization of 77\% for the A exciton was achieved at 90 K with a power density of 0.7 mW/$\mu$m$^2$.
For a quantitative comparison of the $P_C$ values under off-resonance and on-resonance excitations, we performed helicity-resolved PL spectroscopy under both excitation conditions (2.54 eV and 1.96 eV) using identical experimental settings (details provided in SM). Notably, a pronounced negative circular polarization for the IX was observed under resonant conditions (Fig. S15(c) and (f), bottom) compared to the off-resonance condition (Fig. S15(c) and (f), top), attributed to the significantly reduced intralayer valley depolarization \cite{Hsu2018}. To further explore the optical transitions and polarization behavior of both the IX and intralayer excitons in our alloy system, we supplement our experimental findings with a theoretical analysis based on a model bilayer Hamiltonian.
\begin{table*}[t]
\caption{\label{tab:table1}%
Interband transition intensities (Fig. 3(d)).}
\begin{ruledtabular}
\begin{tabular}{cc}
Transition&Intensity\\
\hline
1.\footnote{\label{note1}These transitions are shown in Fig.3.}IL ($\sigma+$)&$|P_{+}(K, \uparrow)|^2 = |\braket{\psi_{CB1, l} |\hat{p}_{+}| K, \uparrow}|^2 = |P_0|^2 (1 - |\alpha_2(E)|^2)$ \\
2.\textsuperscript{\ref{note1}}IX ($\sigma+$)&$|P_{+}(K, \downarrow)|^2 = |\braket{\psi_{CB2, l} |\hat{p}_{+}| K, \downarrow}|^2 = |P_0|^2 (|\alpha_1(E)|^2)$\\
3.  IX ($\sigma-$)&$|P_{-}(K, \uparrow)|^2 = |\braket{\psi_{CB2, u} |\hat{p}_{-}| K, \uparrow}|^2 =|P_0|^2 (|\alpha_2(E)|^2)$\\
4. IL ($\sigma-$)&$|P_{-}(K, \downarrow)|^2 = |\braket{\psi_{CB1, u} |\hat{p}_{-}| K, \downarrow}|^2 =  |P_0|^2 (1 - |\alpha_1(E)|^2)$\\
5.\textsuperscript{\ref{note1}}IL ($\sigma+$)&$|P_{+}(-K, \uparrow)|^2 = |\braket{\psi_{CB1, u} |\hat{p}_{+}| -K, \uparrow}|^2 = |P_0|^2 (1 - |\alpha_1(E)|^2)$\\
6. IX ($\sigma+$)&$|P_{+}(-K, \downarrow)|^2 = |\braket{\psi_{CB2, u} |\hat{p}_{+}| -K, \downarrow}|^2 = |P_0|^2 (|\alpha_2(E)|^2)$\\
7.\textsuperscript{\ref{note1}}IX ($\sigma-$)&$|P_{-}(-K, \uparrow)|^2 = |\braket{\psi_{CB2, l} |\hat{p}_{-}| -K, \uparrow}|^2 = |P_0|^2 (|\alpha_1(E)|^2)$\\
8. IL ($\sigma-$)&$|P_{-}(-K, \downarrow)|^2 = |\braket{\psi_{CB1, l} |\hat{p}_{-}| -K, \downarrow}|^2 = |P_0|^2 (1 - |\alpha_2(E)|^2)$
\end{tabular}
\end{ruledtabular}
\end{table*}
\par
The matrix representation of the bilayer Hamiltonian is shown in SM, Section O, along with the CB and VB eigenvalues and eigenvectors (Eqs. (S2)-(S16)). The resulting energy band diagram is displayed in Fig. 3(d). The diagram shows only the higher energy VB states as their spin-split partners have a much lower energy. On irradiating the bilayer sample with circularly polarized light, the transition matrix element describing the excitation of an electron from the VB to the CB is given by the expression $P_{\pm} = \braket{\psi_{CB} |\hat{p}_{\pm}| \psi_{VB}}$, where $\hat{p}_{\pm} = \hat{p_x} \pm i\hat{p_y}$, $\hat{p_x}$ and $\hat{p_y}$ being the x and y components of the momentum operator. The $+$ ($-$) sign refers to light with helicity $\sigma+$ ($\sigma-)$. The matrix representations of the momentum operator is obtained using the operator identity $ \hat{p}_{\alpha} (\alpha = x, y) = \frac{m}{\hbar} \frac{\partial H}{\partial k_{\alpha}}$, where $m$ is the effective mass of the electron and $H$ the Hamiltonian matrix (Eq. (S2)). After photoexcitation, the number of carriers generated is proportional to the square modulus of the transition matrix element. The dynamics of photocarriers involve three successive steps: the generation of charged photocarriers (holes in the VB and electrons in the CB) in a monolayer through the absorption of light, the relaxation of the carriers towards equilibrium conditions via interlayer charge transfer and finally the recombinations of the electrons and holes resulting in the emission of light detected in PL experiments \cite{PhysRevB.99.195415, PhysRevLett.114.087402}. In Fig. 3(d), the intralayer A exciton is generated from both the VB states, $\ket{K, \uparrow}$ and $\ket{-K, \uparrow}$ through irradiation with light of helicity $\sigma+$. A fraction of the photogenerated electrons in the $-K$  valley relaxes  from the upper layer to the CB minimum in the lower layer of  the same valley, thus forming an IX. 
\par
Table I shows the calculated interband transition intensities for the various allowed transitions in both the $K$ and $-K$ valleys (Fig. 3(d)) with the expressions for $|\alpha_1(E)|^2$ and $|\alpha_2(E)|^2$ given in Eqs. (S15) and (S16) and the suffix $l(u)$ referring to the lower (upper) layer of the bilayer. From the Table, one finds that the emission helicity $\sigma-$ of the IX is opposite to that of the radiation generating the intralayer A exciton, in agreement with our experimental results (Fig. 3(a)-(c)). This feature has been observed previously in specific heterostructures \cite{Hsu2018, doi:10.1126/sciadv.ado1281}. 
The intralayer B exciton is generated in the upper layer of the $–K$ valley through the transition from the spin-split lower VB state (not shown in Fig. 3(d)) to the CB2 state. The reported IX formation, it should be noted, conserves spin. As has been established experimentally, the interlayer charge transfer process is predominantly driven by spin-conserving transfers to the lowest energy band \cite{Hsu2018}. The formation of the IX in the K valley is less favourable as it involves a pair of spin flips.
\par
In TMDC monolayers, a quantitative measure of circular dichroism is provided by the degree of valley polarization, $\rho_{VP}$, i.e., the preferential occupation of the valleys $K$ and $–K$ on photoexcitation with right-circularly-polarized ( helicity $\sigma+$) and left-circularly-polarized (helicity $\sigma-$) light, respectively. The valley-specificity is a consequence of the breaking of the IS in the monolayer. In the case of the pristine 2H-stacked TMDC bilayer, the global IS is not broken though the component monolayers are still inversion-asymmetric. On excitation with laser light of a specific helicity, say, $\sigma+$, both the $K$ and $-K$ valleys are equally occupied (no preferential occupation) so that  $\rho_{VP}$ = 0. In the presence of an electric field (internally generated or external), the global IS is broken and one expects to find a non-zero value for $\rho_{VP}$.
 Considering polarized light with $\sigma+$ helicity, $\rho_{VP}$ is calculated as
\begin{equation}
\rho_{VP} =  \frac{|P_{+}(K,\uparrow)|^2 + |P_{+}(K,\downarrow)|^2 - |P_{+}(-K,\uparrow)|^2 - |P_{+}(-K,\downarrow)|^2}{|P_{+}(K,\uparrow)|^2 + |P_{+}(K,\downarrow)|^2 + |P_{+}(-K,\uparrow)|^2 + |P_{+}(-K,\downarrow)|^2}
\end{equation}
Plugging in the expressions for the intensities from Table I, we get
\begin{equation}
\begin{split}
   \rho_{VP}  = |\alpha_{1}(E)|^2 - |\alpha_{2}(E)|^2\\
    = \frac{\lambda_{v} - Ed/2}{2E_{l}} - \frac{\lambda_{v} + Ed/2}{2E_{u}}
 \end{split}
\end{equation}
where $2\lambda_v$ is the magnitude of the spin-splitting energy in the VB state, $E_l(E_u)$ ) is the
eigenenergy of the VB state when the holes are localized predominantly in the lower (upper)
layer, $E$ is the magnitude of the internally generated electric field and $d$ the separation distance
between the two layers (SM, Section O).
\par
The global IS is restored when the electric field $E$ = 0 so that $E_{u} = E_{l}$ and $\rho_{VP} = 0$.
The inversion-symmetric bilayer has the distinguishing feature of exhibiting circular dichroism based not on valley polarization but on spin polarization arising from an unique interplay between the spin, valley and layer degrees of freedom \cite{PhysRevB.99.195415, PhysRevLett.114.087402} incorporated in the model Hamiltonian (Eq. (S2)). In the case of $WX_2$ bilayers, large $2\lambda$  to $t_{\perp}$ ratios  ($\lambda$ = $\lambda_v$ or $\lambda_c$, is a measure of the strength of the spin-orbit coupling), result in the suppression of  interlayer hoppings for both  electrons and holes \cite{Gong2013, Jones2014, doi:10.1073/pnas.1406960111} giving rise to layer localization in the 2H stacking configuration. Recent
experiments involving spin-, angle- and time-resolved photoemission spectroscopy \cite{PhysRevLett.118.086402, Riley2014, PhysRevLett.117.277201} provide evidence that in centrosymmetric samples
of TMDC bilayers, it is possible to generate spin-, valley- and layer-polarized excited states in
the CB using circularly polarised pump pulses. The degree of spin polarization for light with $\sigma+$ helicity is given by
\begin{equation}
S_{pol} =\frac{|P_{+}(K,\uparrow)|^2 + |P_{+}(-K,\uparrow)|^2 - |P_{+}(K,\downarrow)|^2 - |P_{+}(-K,\downarrow)|^2}{|P_{+}(K,\uparrow)|^2 + |P_{+}(-K,\uparrow)|^2 + |P_{+}(K,\downarrow)|^2 + |P_{+}(-K,\downarrow)|^2}
\end{equation}
From the entries in Table I, we get
\begin{equation}
\begin{split}
    S_{pol} &= 1- |\alpha_{1}(E)|^2 - |\alpha_{2}(E)|^2\\
    & = \frac{\lambda_{v} - Ed/2}{2E_{l}} + \frac{\lambda_{v} + Ed/2}{2E_{u}}
    \end{split}
\end{equation}
We refer to Fig. 3(d) and Table I to understand the origin of valley- and layer-specific out-of-plane spin polarization. Consider the $-K$ valley in which electrons are excited from the VB states
to the CB states using circularly polarized light with helicity $\sigma+$. The two possible transitions
are from the $\ket{-K, \uparrow}$ and $\ket{-K, \downarrow}$ VB states to the upper layer CB1 (spin-up) and CB2 (spin-down) states respectively. The corresponding transition intensities from Table I are $|P_{+}(-K, \uparrow)|^2 = |P_0|^2 (1 - |\alpha_1(E)|^2)$ and $|P_{+}(-K, \downarrow)|^2 = |P_0|^2 (|\alpha_2(E)|^2)$. Since $|P_{+}(-K, \uparrow)|^2 >> |P_{+}(-K, \downarrow)|^2$, $\sigma+$ helicity light predominantly excites up-spin electrons in the upper layer of
the $-K$ valley. Similarly, $\sigma+$ helicity light predominantly excites up-spin electrons in the lower layer of the $K$ valley. The chain of arguments can be extended to the case of $\sigma-$ helicity light
demonstrating spin-, valley- and layer-specific excited states in the CB.

In the case of the tungsten-based bilayers, $t_{\perp}$ is $<<$ $\lambda_{v}$, so that $S_{pol}$ is close to unity. Du et al. \cite{PhysRevB.99.195415} have defined the degree of CP as
\begin{equation}
    P_C= \frac{N_{+} - N_{-}}{N_{+} + N_{-}}
\end{equation}
where $N_{+ (-)}$ is the total number of photocarriers associated with $\sigma+(\sigma-)$  helicity of light, taking into account the relaxation of photogenerated carriers.  In a PL experiment, since the measured intensity is proportional to the number of carriers, $P_C$ is given by
\begin{equation}
    P_C = \frac{I(\sigma+)-I(\sigma-)}{I(\sigma+)+I(\sigma-)}
\end{equation}
where $I(\sigma+)$ and $I(\sigma-)$ are the right-CP and left-CP luminescence, respectively. The theoretically calculated expression is given by \cite{PhysRevB.99.195415}
\begin{equation}
    P_C= \frac{1}{2}\left[\frac{(\lambda_{v} + Ed/2)^2}{(\lambda_{v} + Ed/2)^2 + |t_{\perp}|^2} + \frac{(\lambda_{v} - Ed/2)^2}{(\lambda_{v} - Ed/2)^2 + |t_{\perp}|^2}\right]
\end{equation}
When $E$=0, one gets  $P_C=(S_{pol})^2$ showing that both the polarization measures are linked to spin polarization.
\subsubsection{Theory and experiment-based estimates}
An estimate for $Ed$ can be obtained by matching our experimental data with theoretical results. The peak positions of the A, B excitons, and the IX, as determined from PL measurements, are 1.877 eV, 2.263 eV and 1.710 eV respectively. The energy difference between A and B is given by 0.386 eV. From the bilayer model, the energy difference between A and B excitons is given by $2\lambda_v - 2\lambda_c = 0.3969$ eV ($2\lambda_v$ = 0.4252 eV, $2\lambda_c$ = 0.02832 eV for the alloy bilayer, see SM, Section O), in close agreement with the experimental result. The difference in the peak positions of the A exciton and IX is $2\lambda_c + Ed$ = 0.167 eV, yielding an estimate of $Ed$ = 0.14 eV.
\par
With $t_{\perp} \sim $ 0.056 eV, close to the value for bilayer WS$_2$ (the alloy material is sulfur rich), the degree of valley polarization, as given by Eq. (2), is $\rho_{VP}$ = -0.024, i.e., about  - 2\%, a very small value. 
From Eq. (4), the degree of spin polarization is $S_{pol}$ = 0.957.  From Eq. (7), the degree of circular polarization, $P_C$ = 0.92, i.e., 92\%. In arriving at this estimate, depolarizing mechanisms like phonon-assisted intervalley scattering have not been taken into account which would further reduce the value of $P_C$. For our alloy sample, the highest value of the degree of circular polarization that could be achieved is about 77\% at 90 K. 
\subsubsection{Temperature dependence of PL and degree of circular polarization}
The homobilayers WS$_2$, WSe$_2$, as well as the bilayer alloy material are indirect bandgap semiconductors in which both direct and indirect optical transitions contribute to the PL emission. The computed electronic band structures of these materials show that the prominent valleys of interest are the $K$, $\Lambda$ and $\Gamma$ valleys \cite{Ernandes2021, Godiksen, Lindlau2018}. Though the indirect transitions across the valleys are typically unpolarized, their influence on the degree of circular polarization of the direct $K$-$K$ transition is apparent in the temperature dependence of the polarization degree. As temperature changes, the nature of the dominant indirect transition channel may also change due to the shifts in energy of the valley minima and maxima. The polarization of the direct $K$-$K$ transition depends on the energy difference of the CB minima at the $K$ and $\Lambda$ points. The $\Lambda$ valley is located in-between the $K$ and $\Gamma$ valleys and has been shown to play a critical role in modulating the temperature dependence of the PL emission intensity and polarization of the bright neutral (A) exciton \cite{Godiksen}. In the case of the WSe$_2$ bilayer, an indirect band gap semiconductor, the dominant indirect transition is $\Lambda$-$\Gamma$ ($K$-$\Gamma$) below (above) a crossover temperature. As observed in polarization- and temperature-dependent PL experiments \cite{Godiksen}, the DOCP of the $K$-$K$ transition increases as temperature decreases below the crossover point. Above this temperature, the $K$-$\Gamma$ indirect transition becomes more favourable resulting in a reduced polarization of the $K$-$K$ transition due to the depletion of electrons from the K valley. At room temperature, the DOCP of the transition has a negligibly small value. On the other hand, in the case of the WS$_2$ bilayer, the $\Lambda$-$\Gamma$ indirect transition is the dominant one over the whole temperature range and the DOCP of the $K$-$K$ transition has a finite value even at the room temperature. In each case, the DOCP decreases as temperature increases. Here, we point out that the alloy bilayer, WS$_{2x}$Se$_{2(1-x)}$, reduces to the compositions of the pristine bilayers, WSe$_2$ and WS$_2$, for $x=0$ and $1$ respectively. In the case of our sample, $x=0.88$, showing its proximity compositionally to the WS$_2$ bilayer, with  both the bilayers exhibiting a high DOCP value for the $K$-$K$ transition (A exciton) at low temperatures.
The intralayer exciton states can be classified as optically bright and dark exciton states. A radiative recombination is optically forbidden for a dark exciton state due to the mismatch of either the momentum or the spin between the VB and CB states.
\par
In the light of the above comments, the results, displayed in Figs. 4(a)-(d), can be understood in the following manner. In the monolayer limit, the tungsten-based TMDC materials, (WX$_2$, X = S, Se) have the dark A exciton state (spin-forbidden) as the ground state. Due to the presence of the dark state, the PL intensity is quenched at low temperatures. A rise in the temperature brings about thermal activation from the dark to the bright A-exciton state thus increasing the emission intensity in the PL spectrum. The enhancement of the PL intensity of the A exciton as a function of temperature has been experimentally confirmed in monolayer WSe$_2$ \cite{PhysRevLett.115.257403}. On the other hand, in our case of the tungsten-based bilayer alloy material, the emission intensity of the A exciton decreases as the temperature increases. At a raised temperature, one expects the number of electrons in the bright A exciton state to increase due to thermal excitation. The number is, however, depleted due to a number of electron transfer processes. On an ultrafast timescale, a fraction of the electrons in one layer is transferred to the other layer in the same valley forming thereby the IX. This time scale of the interlayer charge transfer is much shorter than the recombination time of the electron and hole in the A exciton. In a heterostructure, ultrafast interlayer charge hopping has been shown to strongly quench the intralayer exciton PL \cite{Hong2014}. Moreover, indirect electron transitions to different valleys are facilitated by enhanced phonon/defect scatterings at increased temperatures. In fact, efficient intervalley scattering has been suggested to be the major factor for the strong
quenching of the PL signal in multilayer TMDC samples \cite{PhysRevLett.117.277201}. These processes result in the PL intensity of the A exciton decreasing as the temperature increases (Fig. 4(b)) with a consequent reduction in the degree of circular polarization (Fig. 4(d)).
\par
Although the exciton landscape of the bilayer alloy material describing a variety of intralayer and interlayer bright and dark exciton states is not known in detail, it is marked by the presence of low-lying dark exciton states, both spin and momentum-forbidden \cite{Godiksen}. As in the case of the WSe$_2$ monolayer, the dark exciton states require thermal activation to the bright exciton states in order that optically-allowed PL emission is possible. In Fig. 4(b), the temperature dependence of the PL intensity of the IX is fitted by the Boltzmann distribution, arising from a simplified two-level model in which a population of bright states is separated by energy $\Delta E$ from a population of low-lying dark states \cite{PhysRevLett.115.257403, Godiksen}. The fitting equation is given by
\begin{equation}
    I_{PL}(T) = A \frac{exp\left(\frac{-\Delta E}{k_{B}T}\right)}{1 + exp\left(\frac{-\Delta E}{k_{B}T}\right)}
\end{equation}
where $I_{PL}(T)$ is the measured PL intensity at temperature T, A is a constant factor and $k_{B}$ is the Boltzmann constant. From the fit, we obtain $\Delta E$ = 20 meV. The fitting indicates that, due to the presence of lower-lying dark states, the intensity of the IX emission decreases as temperature is lowered. In the expression for the degree of circular polarization, $P_C$ (Eq. (6)), the PL intensity of the IX peak is given by the cross-polarized component $I(\sigma-)$. Around the peak energy of the IX, $I(\sigma-)$ dominates over the co-polarized component $I(\sigma+)$ so that the degree of polarization is negative (Fig. 3(c)). The temperature dependence of the dominant component,  $I(\sigma-)$, suggests that the degree of polarization becomes more negative as temperature goes up (Fig. 4(d)). This type of temperature-dependent PL intensity behavior has not been reported earlier except in the case of a heterostructure comprising three layers of WSe$_2$ and a monolayer of MoS$_2$, suggested to be an indirect bandgap semiconductor with low-lying dark exciton states \cite{doi:10.1126/sciadv.abh0863}. The data in Fig. 4(c) are well-fitted by the Varshni equation (Section L in SM). In the case of the A exciton, the excitonic band gap decreases as the temperature increases so that the peak energy of the PL is red shifted (Fig. 4(c)). On the other hand, the blue-shifting of the IX peak, as depicted in Fig. 4(c), is attributed to the repulsive dipole-dipole interactions among the IXs \cite{Jiang2021, Rivera2015}. As the temperature rises, the population density of IX excitons increases, resulting in an enhanced PL emission intensity (Fig. 4(b)) and also the blue-shifting of the IX peak due to the strengthened repulsive interactions among the more numerous IXs. In contrast to our observation, the IX exciton peak shifts to higher energies as the temperature decreases in heterostructures like MoS$_2$/MoSe$_2$/MoS$_2$ \cite{doi:10.1021/acs.nanolett.7b03184}. The difference in the observations could possibly arise from the distinctive electronic band structures of tungsten and molybdenum-based TMDC materials. In SM, Section P, we include a data analysis of the PL spectrum at 90 K suggesting a double-peaked structure of the IX, in conformity with recent experimental observations \cite{Yu_2018, Yu:20, Durmuş2024}. Additional measurements are required for detailed analysis of this observation.
\par
We interprete the temperature dependence of the DOCP of the A exciton  (Fig. 4(d)) as follows. The steady state degree of circular polarization (DOCP), $P_C$, is given by the expression,
\begin{equation}
    P_C= \frac{P_0}{1+2\frac{\tau}{\tau_s}}
\end{equation}
where $P_0$ is the theoretical limit of the PL polarization and $\tau$, $\tau_s$ denote the exciton lifetime and the spin polarization lifetime of the electrons. The spin polarization, rather than the valley polarization, is measured in the polarization-resolved PL experiments in the cases of both the pristine/gated homobilayers, WS$_2$, WSe$_2$ and our alloy bilayer sample. For the A exciton generated by $\sigma+$ helicity light excitation, the intensity difference between the $\sigma+$ and $\sigma-$ components of the PL signal provides an estimate of the population imbalance between the spin-up and spin-down excitons with the $\sigma+$ helicity light exciting spin-up electrons at both the $K$ and $-K$ valleys. The spin-down population owes its origin to the depolarization of the spin-up population. The two lifetimes, $\tau$ and $\tau_s$, depend on the depolarization mechanisms specific to the bilayer. A detailed discussion of these mechanisms is given in Refs. \cite{doi:10.1073/pnas.1406960111, Godiksen, Niu_2018}. The lifetimes can be measured in experiments involving the time-resolved PL (TRPL) and time-resolved Kerr rotation (TRKR) techniques, which probe the  exciton depolarization dynamics. From Eq. (9), it is clear that to achieve a high value of the DOCP, $\tau_s$ should be considerably larger than $\tau$. As shown by Niu et al. \cite{ Niu_2018}, in the case of the WSe$_2$ bilayer, the decay process of the electron spin polarization, as probed by TRKR, is temperature-dependent and the decay process is well-fitted by a biexponential decay curve with a short and a long decay time. At T=10 K, these times are $\sim$ 3.8 ps and $\sim$ 20 ps respectively. The short polarization decay time is a measure of the exciton lifetime $\tau$. The long decay time, a measure of $\tau_s$, arises from the rapid intervalley scattering from the $K$ valley to the $\Lambda$ valley (a global minimum) followed by a rapid interlayer charge transfer. Both the intervalley scattering and interlayer charge transfer are enhanced at higher temperatures giving rise to diminished values of $\tau_s$. Assuming that the ratio $\frac{\tau}{\tau_s}$ follows a Boltzmann distribution, $c.exp(-\Delta E/ k_B T)$, where $c$ is a constant and $\Delta E$ is the activation energy needed for the intervalley scattering \cite{Godiksen}, the DOCP as a function of temperature is obtained from Eq. (9). This functional dependence is fitted to the experimental data for the A exciton (green curve in Fig. 4(d)). The values of the fitted parameters are: $c=5.25$, $\Delta E=67$ meV. From our experimentally measured value of the DOCP of the A exciton, 0.77 at  90 K, the value of  $\tau/ \tau_s$ $\sim$ 0.15 from Eq. (9). The same value is $\sim$ 0.19 at 10 K in the case of the WSe$_2$ bilayer.
\par
As pointed out in \cite{doi:10.1073/pnas.1406960111}, a number of factors may collectively contribute to
the robust circular polarization of bilayer materials. These include shorter
exciton lifetime, smaller exciton binding energy, extra spin-conserving
channels via intervalley-interlayer scatterings, absent in monolayers, and
lastly, the spin-valley-layer coupling characterising bilayers in the 2H
stacking configuration, which reduces intervalley scattering. Comprehensive
quantitative studies are required for the elucidation of the different
mechanisms contributing to the robust circular polarization of the bilayer.
\subsection*{Discussion}
In our bilayer alloy sample, changing the proportion of the chalcogen atoms S and Se allows for the tuning of both the band gap as well as the magnitude of the net out-of-plane electric field. In the case of the Janus bilayer material, WSSe, one chalcogen layer consists solely of S and the other layer of Se atoms. In this case, the magnitude of the internal electric field is expected to be the largest. The generation of IX excitons in the Janus bilayer has, however, not been demonstrated as yet. While disorder in the occupation of the chalcogen layers in our alloy sample has an indirect effect on the type-II band alignment, via the generation of the internal electric field, a simple bilayer model, focused on the central layer of tungsten atoms, suffices to capture the essential features of our experimental observations. As in the case of TMDC monolayers, the dominant contributions to the bilayer physics mostly come from the $d$-orbitals of the tungsten atoms. A new effect termed “hidden spin polarization” has been uncovered in recent spin- and angle-resolved photoemission spectroscopy (spin-ARPES) experiments on the pristine bilayer materials WSe$_2$ and MoS$_2$ \cite{Riley2014, PhysRevLett.118.086402, PhysRevLett.117.277201}, opening up the exciting prospect of the generation and manipulation of spin polarization in a centrosymmetric, nonmagnetic material. The “hidden spin polarization” refers to the situation in which global IS forbids the appearance of spin polarization globally whereas, locally, spin polarization exists in the inversion asymmetric monolayers, detected through spin-ARPES experiments with selective focus on single layers in specific valleys.
\par
In summary, the disorder in the chalcogen layers in our bilayer alloy sample indirectly influences the type-II band alignment by introducing an internal electric field. This alignment facilitates the formation of interlayer excitons in the alloy bilayer sample, thus providing a new route for the generation of IX excitons. While PFM studies provide the direct evidence of the internal electric field, laser-power, temperature, gate voltage, and polarization-dependent PL measurements provide confirmatory evidence for the existence of IX excitons. Combining experimentally observed features with the theoretical analysis of a toy bilayer model, our study highlights a unique interplay between spin, valley and layer degrees of freedom. Our findings offer valuable insights into excitonic behaviour in bilayer systems, paving the way for advancements in spintronic and optoelectronic applications.

\newpage
\section*{Supplemental Material}
\vspace{0.5 cm}
\hrule
\vspace{1 cm}
\setcounter{figure}{0}
\setcounter{equation}{0}
\renewcommand{\thetable}{S\arabic{table}}
\renewcommand{\tablename}{Table}
\renewcommand{\thefigure}{S\arabic{figure}}
\renewcommand{\figurename}{Fig.}
\renewcommand{\theequation}{S\arabic{equation}}

\subsection{CVD synthesis of TMDC alloy WSSe}
\begin{figure*}[h!]
 \centering
\includegraphics[width=0.8\linewidth]{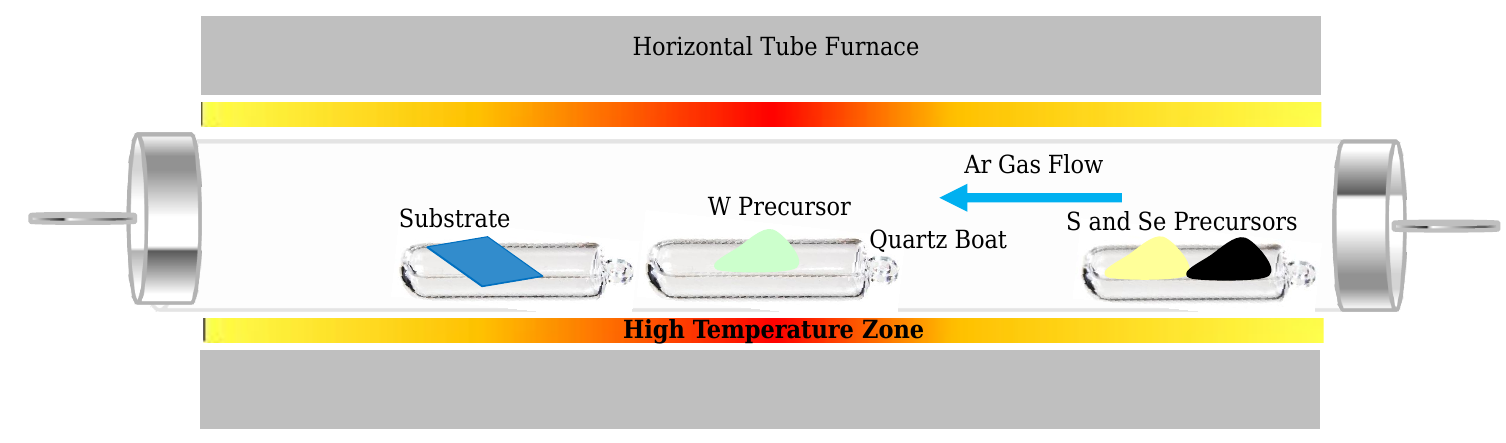}
\caption{Schematic representation of the sample preparation process using the CVD technique.}
\end{figure*}
TMDC alloy WSSe has been synthesized on a SiO$_2$/Si substrate using the chemical vapor deposition (CVD) technique in a homemade horizontal tube furnace equipped with an alumina tube (45 mm diameter) and vacuum pump. Figure S1 illustrates a schematic representation of the CVD setup. For this growth, tungsten(VI) oxide and sulfur powders were purchased from Sigma-Aldrich. Selenium powder was taken from Alfa Aesar. During the synthesis process, optimization of various parameters, including the distances between the W, S, and Se precursors, temperature, and flow rates, was carried out carefully. A quartz boat containing WO$_3$ powder was positioned within the tube in a high-temperature zone ($\sim$950 $^{\circ}$C), while the S and Se powders were placed in a relatively low-temperature zone ($\sim$200$^{\circ}$C). Before the chemical reaction, the tube was evacuated by a rotary pump and purged several times with high-purity Ar gas to remove oxygen and other contaminants. The temperature of the furnace was gradually increased at a rate of 25 $^{\circ}$C/min with a flow of 10 sccm of high-purity Ar gas, reaching its maximum temperature over a span of 30 minutes. The furnace was maintained at the maximum temperature for 10 minutes with 40 sccm Ar flow before being allowed to cool down naturally to room temperature.
\subsection{Experimental tools}
The compositional analysis was performed using a ZEISS scanning electron microscope equipped with an energy-dispersive X-ray (EDX) spectrometer. PFM measurements were conducted (in contact mode) using an AFM equipped with a dual AC resonance tracking piezoresponse module (MFP-3D$^{TM}$, Asylum Research). Room-temperature optical measurements, such as photoluminescence (PL) and Raman spectroscopy, were conducted using a micro-Raman spectrometer (LabRAM HR, Horiba Jobin Yvon) in back-scattering geometry, coupled with a Peltier-cooled CCD detector. Excitations at two different energies were employed: one corresponding to an air-cooled argon-ion laser with an energy of 2.54 eV (wavelength of 488 nm), and the other from a HeNe laser with an energy of 1.96 eV (wavelength of 633 nm). The excitation light was focused onto the sample using a 50x objective lens with a numerical aperture (NA) of 0.75. For the Raman measurements, a grating with 1800 grooves/mm was used, whereas a grating with 600 grooves/mm was utilized for the PL measurements. Temperature-dependent PL measurements were conducted using a cryostat (Linkam THMS600) with a continuous flow of liquid nitrogen to control the temperature. For linear polarization dependent measurement, we used vertically polarized light to excite the sample and collect the data
in both vertical (co-polarized) and horizontal (cross-polarized) configurations using an analyzer. For circular polarization-resolved PL measurements, the polarization state of both excitation and detection beams were controlled using a combination of linear polarizers and a quarter-wave plate ($\lambda$/4) installed in the excitation and detection paths.
\par
To define the electrical contacts for gate voltage-dependent PL studies, the samples were first subjected to standard solvent cleaning using acetone and isopropanol (IPA), followed by nitrogen blow-drying and brief thermal treatment on a hotplate to eliminate residual moisture. A uniform layer of AZ1512-HS positive photoresist was then applied via spin coating and soft-baked at 100 $^0$C for 1 minute. Suitable flakes were identified on SiO$_2$/Si substrate under an optical microscope, and the contact pattern was written using a laser lithography system (Microtech LW405) with an exposure dose of 160 mJ/cm$^2$. Subsequently, a bilayer of Cr/Au (5 nm/60 nm) was deposited via electron-beam evaporation. Lift-off of the photoresist was performed in acetone to reveal the patterned metal contacts. Finally, electrical connections were established by wire bonding using gold wire and silver epoxy. The SiO$_2$ layer beneath the sample serves as an insulator, separating it from the doped silicon back-gate.
\subsection{Basic characterizations}
\begin{figure*}[h!]
 \centering
\includegraphics[width=0.8\linewidth]{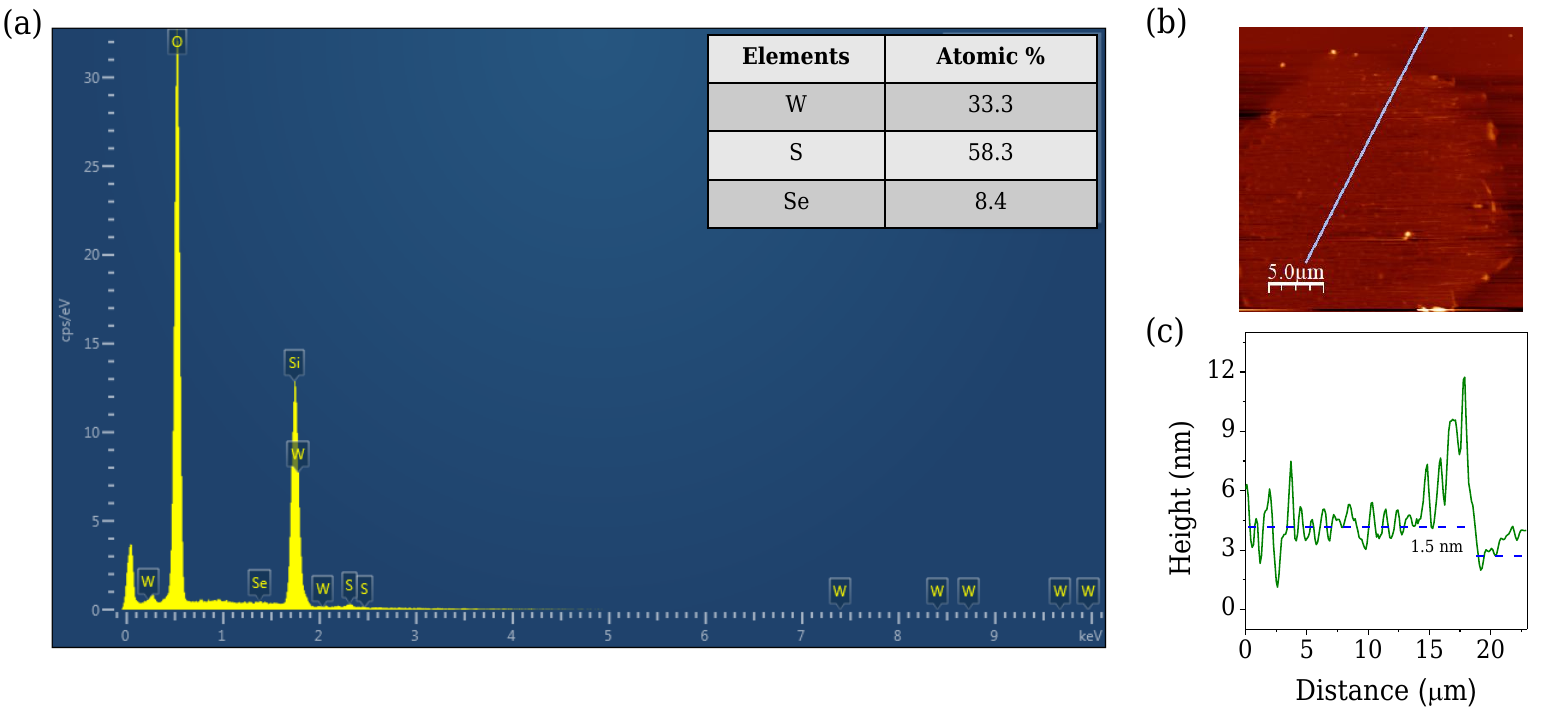}
\caption{Basic characterizations of the as-synthesized WSSe alloy: (a) EDX spectrum (the elemental percentages of W, S, and Se in the WSSe alloy sample are provided in the inset table). (b) AFM topography and (b) height profile from AFM analysis of as-synthesized WSSe sample.}
\end{figure*}
\subsection{Power dependent PL at room temperature}
\begin{figure*}[h!]
 \centering
\includegraphics[width=0.3\linewidth]{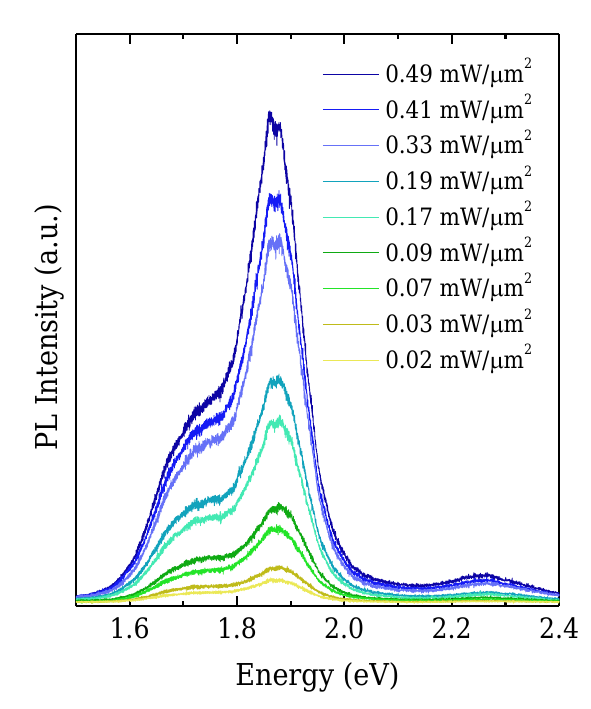}
\caption{PL spectra of bilayer WSSe alloy recorded at room temperature (RT) for different excitation powers, using a 2.54 eV excitation.}
\end{figure*}
\newpage
\subsection{Homogeneity study}
\begin{figure*}[h!]
 \centering
\includegraphics[width=0.6\linewidth]{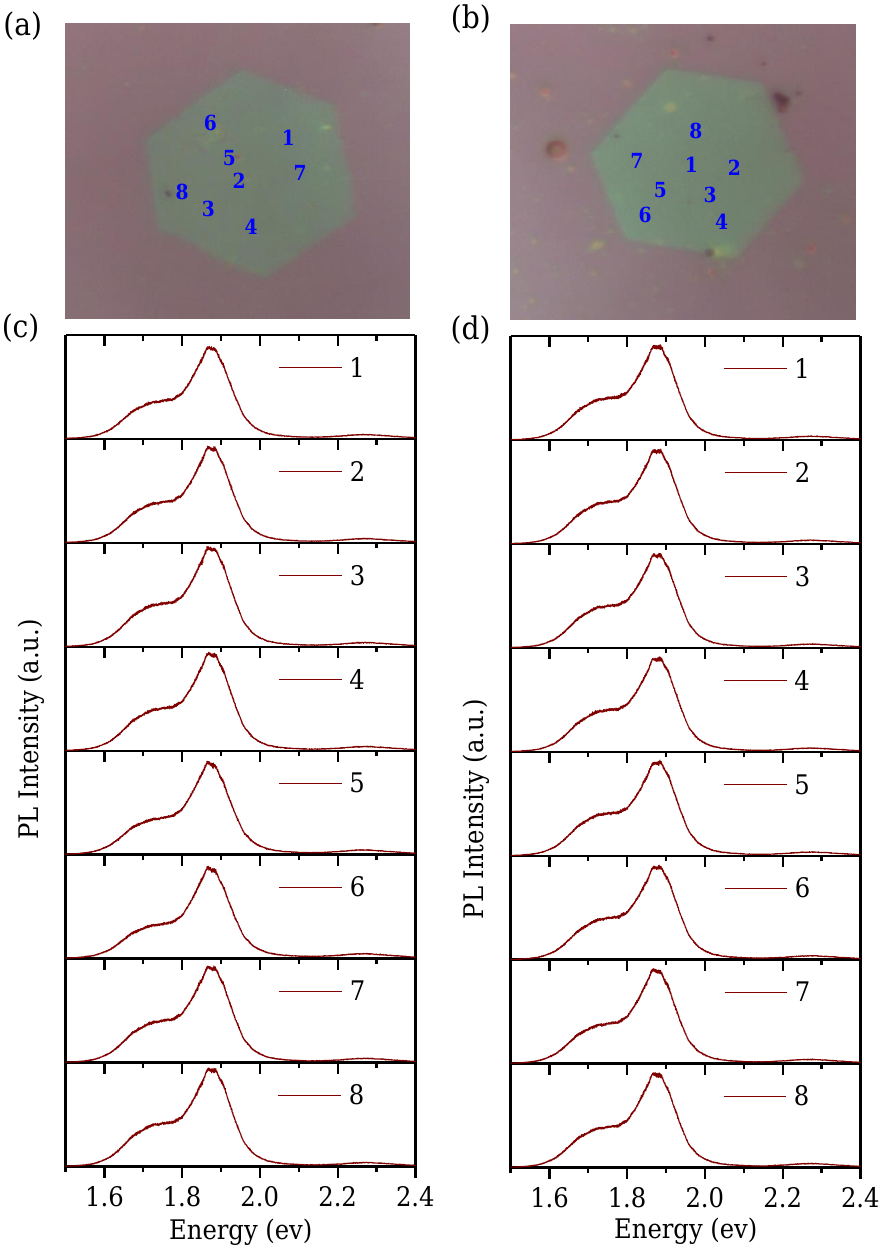}
\caption{PL spectra recorded from different positions across two additional samples (16 positions in total), demonstrating the homogeneity of our CVD-grown sample. Measurements were performed at room temperature using 488 nm excitation.}
\end{figure*}
\newpage
\subsection{Power-dependent PL of interlayer exciton at 90 K}
\begin{figure*}[h!]
 \centering
\includegraphics[width=0.3\linewidth]{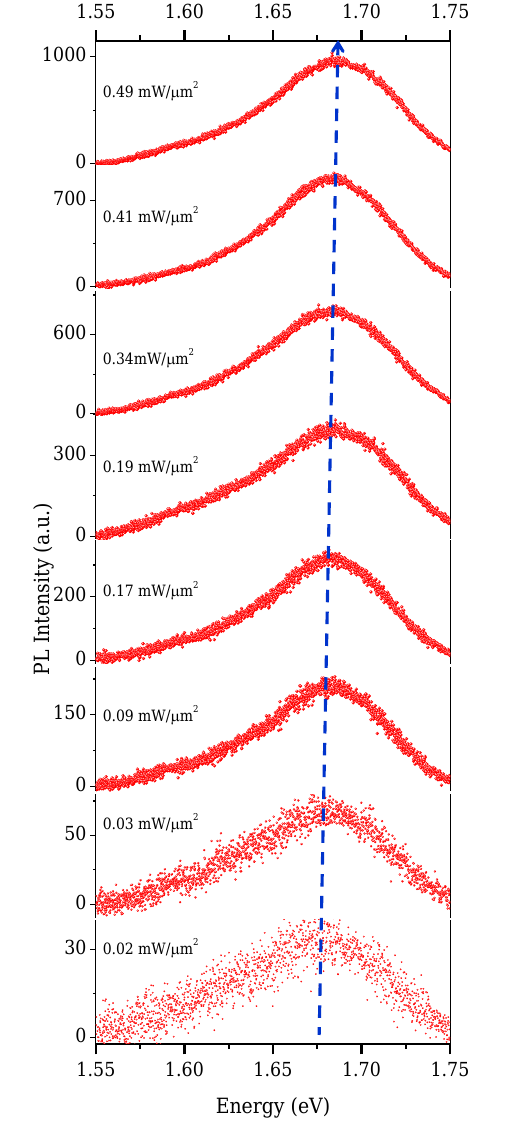}
\caption{Power dependent IX emission at 90 K.}
\end{figure*}
\subsection{Gate-dependent PL study}
 \begin{figure*}[h!]
 \centering
\includegraphics[width=0.55\linewidth]{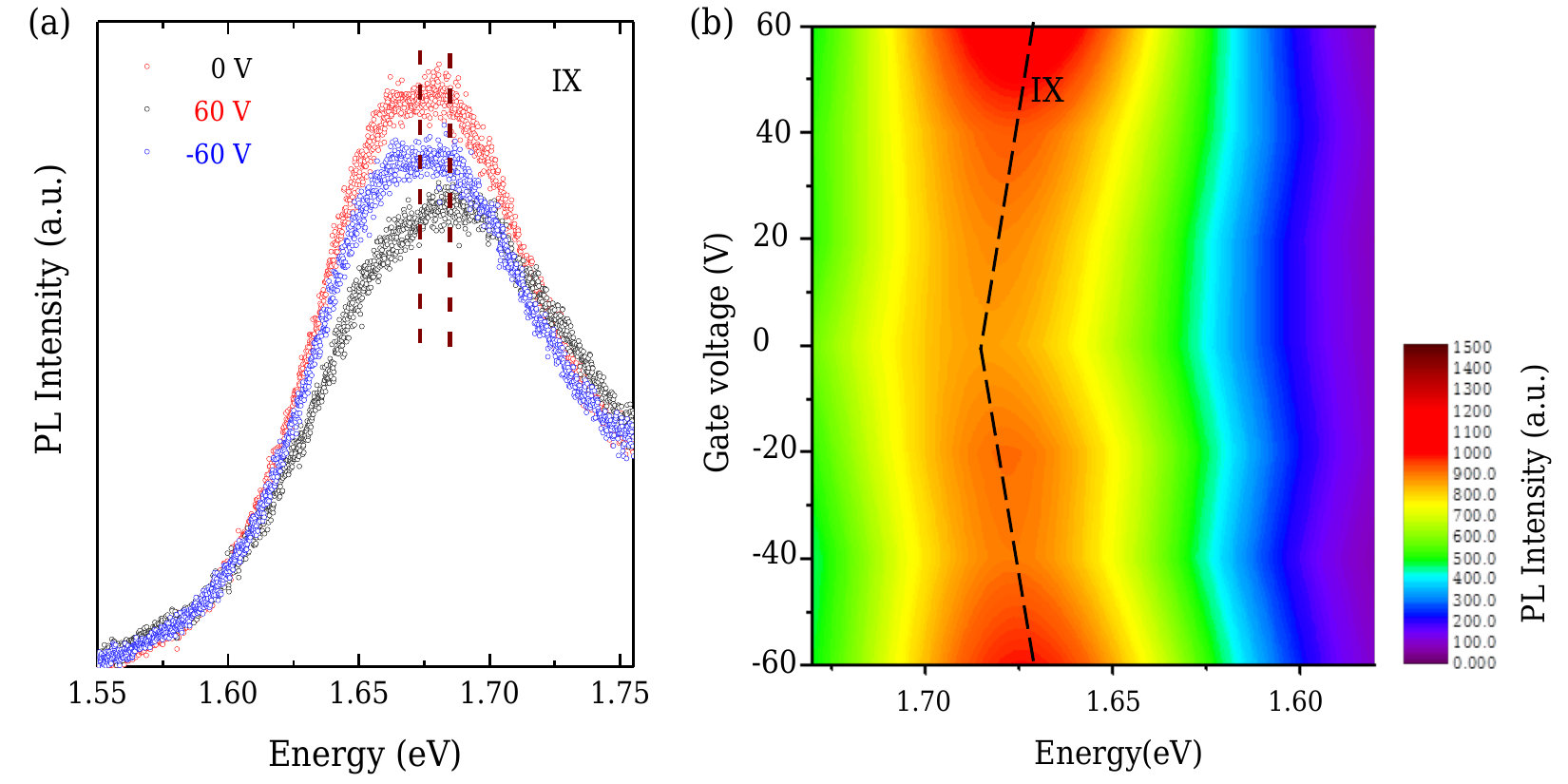}
\caption{Gate-dependent IX emission at 133 K. (a) IX PL spectra recorded at back-gate voltages of 0 V, +60 V, and –60 V. (b) Colour-map showing the evolution of the IX emission as a function of gate voltage. The shift in IX peak energy is traced by the black dotted line.}
\end{figure*}

\subsection{Linear-polarization dependent PL spectra at room temperature}
The response of the A exciton and interlayer exciton (IX) peaks to linearly polarized excitation is shown in Fig. S7 (a for 488 nm excitation and b for 633 nm). A linearly polarized light can be treated as a coherent
superposition of $\sigma+$ and $\sigma-$ helicity light with a certain phase difference, which
determines the polarization direction. From the PL spectrum, the degree of linear polarization can be expressed as, $P_{LP}= ({I_{parallel}-I_{perpendicular}})/({I_{parallel}+I_{perpandicular}})$, where the $I_{parallel}$ ($I_{perpendicular}$) corresponds the intensity of PL with parallel (perpendicular)
polarization with respect to the excitation polarization. The degree of linear polarization is a measure of valley coherence,
i.e., the optical generation of a quantum coherent superposition of exciton states in
the $K$ and $-K$ valleys. The ability to control and manipulate valley coherence has
significant application potential in quantum technologies, specifically, in quantum
computation and information processing. 
\par
The degree of linear polarization ($P_{LP}$) for the A exciton and the IX exciton in our bilayer alloy sample is found to be 52\% and 7.5\%, respectively, at room temperature under near on-resonance excitation (633 nm) with the A exciton. As a reference, in a WS\(_2\) bilayer, the \( P_{LP} \) of the A exciton decreases from 80\% at 10 K to 50\% at room temperature \cite{doi:10.1073/pnas.1406960111}.
\begin{figure*}[h!]
 \centering
\includegraphics[width=0.6\linewidth]{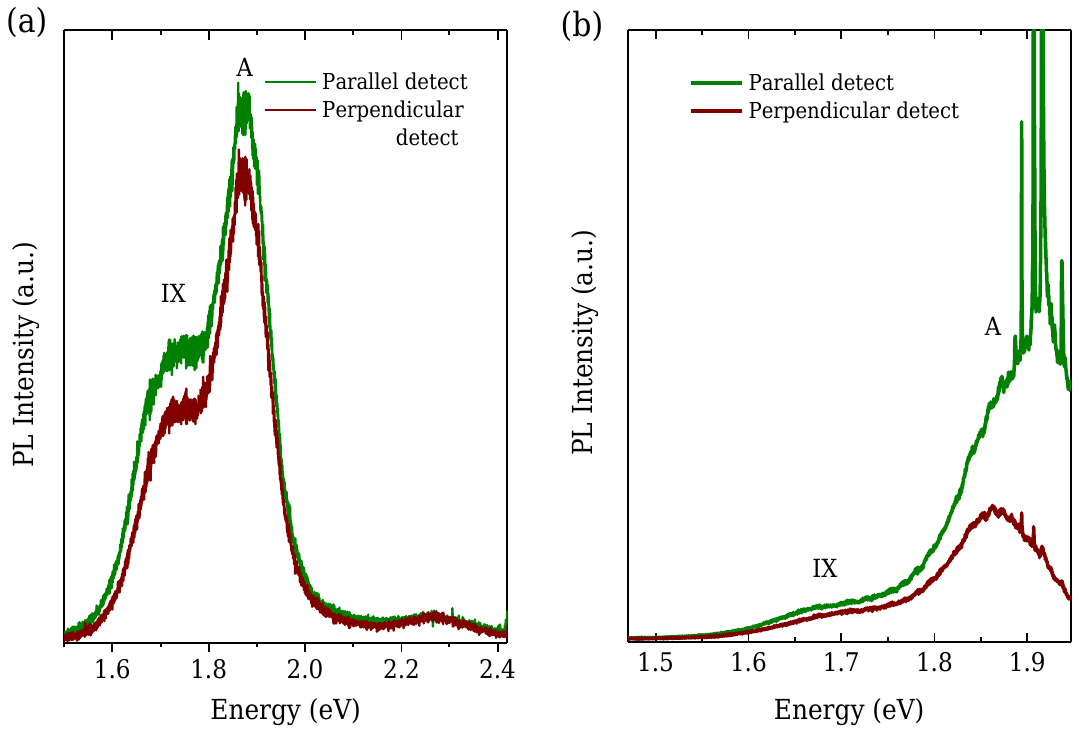}
\caption{Linear-polarization dependent PL spectra at room temperature: (a) for 488 nm and (b) 633 nm (near on-resonance with A exciton) excitation.}
\end{figure*}
\subsection{Piezoelectric force microscopy (PFM)}
\begin{figure*}[h!]
 \centering
\includegraphics[width=0.7\linewidth]{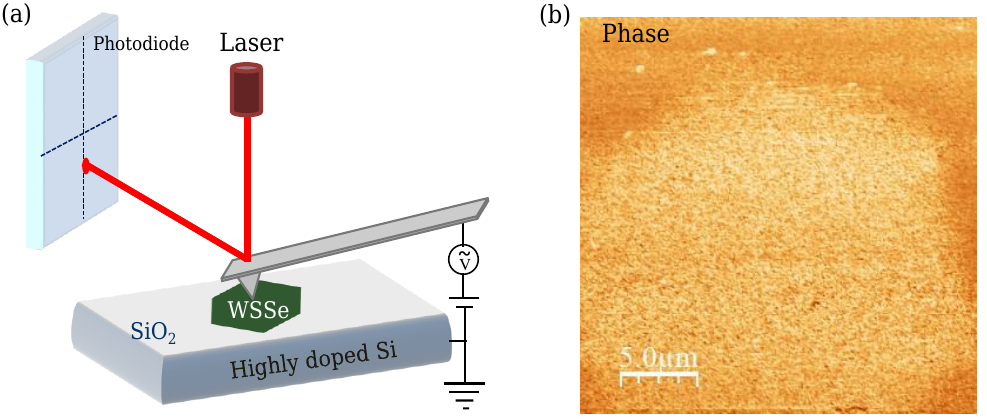}
\caption{PFM study. (a) Schematic diagram of the PFM measurement setup and (b) piezoelectric phase of bilayer WSSe on SiO$_2$/Si substrate.}
\end{figure*}

To investigate the presence of an out-of-plane electric field (or dipole moment) caused by structural asymmetry in our TMDC alloy WSSe, we performed PFM measurements. 
The PFM setup is schematically shown in Fig. S8(a). The voltage applied to the AFM cantilever tip in contact with the sample probes the electromechanical interaction between the tip and the sample and consists of both DC and AC
components, $V_{tip} = V_{DC} + V_{AC}$ cos$(\omega t)$, where $\omega$ is the AC drive frequency. Since our focus is on detecting the out-of-plane dipole moment, the piezoresponse was measured in vertical PFM mode, which captures sample strain in the $z$ direction. The sample to be investigated under the tip experiences local surface deformations, either expansion or contraction, depending on whether the local sample polarization associated with the vertical dipole moment, aligns parallel or anti-parallel to the applied AC field. The periodic AC voltage causes the surface to oscillate, which bends the cantilever periodically. This deflection is detected as a change in the laser spot position on a photodetector. The local piezoelectric response is extracted as the first harmonic component, $A_{1\omega}$, of the tip deflection, $A = A_0 + A_{1 \omega} cos(\omega t + \phi)$, using a lock-in amplifier, allowing sub-nanometer deformations to be distinguished from nanometer-scale topography features. The signal is characterized by amplitude $A_{1\omega}$ (in nm), which reflects electromechanical coupling strength, and phase $\phi$, which indicates polarization direction of the sample. The phase $\phi = 0^{\circ}$  (or 180$^{\circ}$) corresponds to polarization parallel (or anti-parallel) to the AC field, meaning the surface oscillations are in-phase (or out-of-phase) with the voltage oscillations. To enhance the signal-to-noise ratio, piezoelectric measurements were conducted using an AFM equipped with a dual AC resonance tracking piezoresponse module (MFP-3D, Asylum Research). At the contact resonance frequency, cantilever oscillation reaches its maximum amplitude. The cantilever response is amplified by the cantilever's quality factor ($Q$ $\sim$ 10-100).
In PFM domain imaging, $V_{DC}$ is typically set to zero, while an AC voltage applied to the cantilever tip scans the sample, producing piezoresponse amplitude and phase images. These images reveal the location and polarity of the polarization domains in the sample. Figures 2(a) and S8(b) display the amplitude domain and phase images, respectively. Further analysis includes amplitude (strain) $A_{1\omega}$ versus $V_{DC}$ and phase versus $V_{DC}$  plots, measured in the presence of an AC probing voltage, as discussed in the main paper. 
\par
In the amplitude versus DC voltage plot, a butterfly-shaped loop is observed, representing the strain–electric field hysteresis loop, which originates from the inverse piezoelectric effect. This behavior has been extensively described by D. Damjanovic \cite{Damjanovic1998}, outlining the changes along the loop's branches in an ideal scenario. For a polarized domain, when the electric field is aligned either parallel or antiparallel to the polarization, the sample expands or contracts, respectively \cite{Damjanovic1998, PhysRevB.109.115304}.
Additionally, PFM measurements on the SiO$_2$/Si substrate are presented in Fig. S9(a) and (b). These results do not exhibit any characteristic piezoresponse, such as polarization switching, indicating that the observed piezoelectric behavior (shown in main text) originates exclusively from the bilayer WSSe sample. The piezoelectric coefficient $d_{33}$ for the alloy material is estimated using the relation linking the strain amplitude to the AC voltage, $A_{1\omega} = d_{33}V_{AC}Q$, yielding a value of $d_{33}$ = 3.4 pm/V for $Q$ = 100. It is important to note that the $d_{33}$ value is qualitative, as small variations in the electrical properties of the sample may affect the piezoelectric response.
\begin{figure*}[h!]
 \centering
\includegraphics[width=0.7\linewidth]{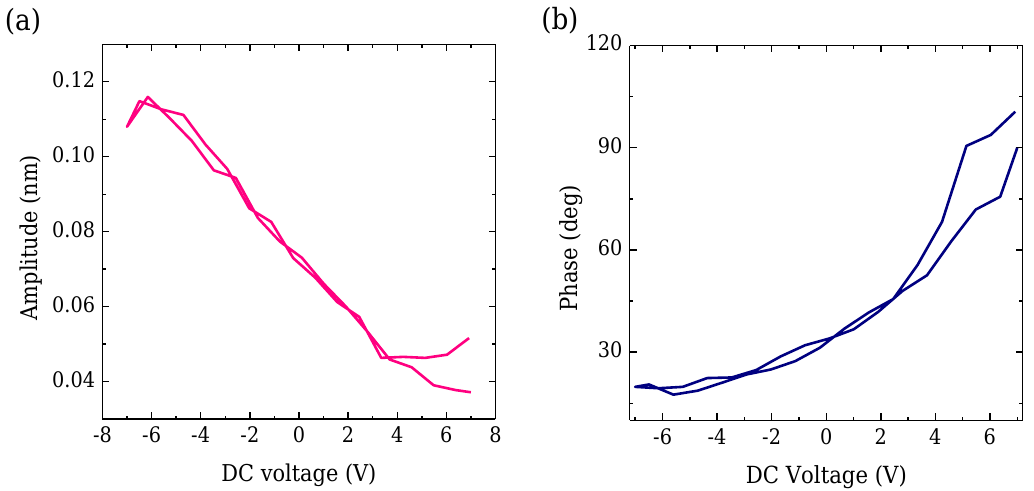}
\caption{PFM measurements on a SiO$_2$/Si substrate. (a) PFM amplitude and (b) phase as a function of DC voltage.}
\end{figure*}
\newpage
\subsection{Helicity-resolved PL measurements using $\sigma-$ excitation at 2.54 eV}
We also performed the helicity-resolved PL measurements using $\sigma-$ excitation of laser energy 2.54 eV, with $\sigma-$ and $\sigma+$ detection, as shown the experimental results in Fig. S10(a). It is observed that the intralayer excitons exhibit a small but positive degree of polarization. Notably, the IX displays circularly polarized emission with helicity ($\sigma+$) opposite to that of the optical excitation ($\sigma-$), resulting in a negative degree of circular polarization (Fig. S10(b)), consistent with the measurements conducted using $\sigma+$ excitation (Fig. 3(a)).
\begin{figure*}[h!]
 \centering
\includegraphics[width=0.6\linewidth]{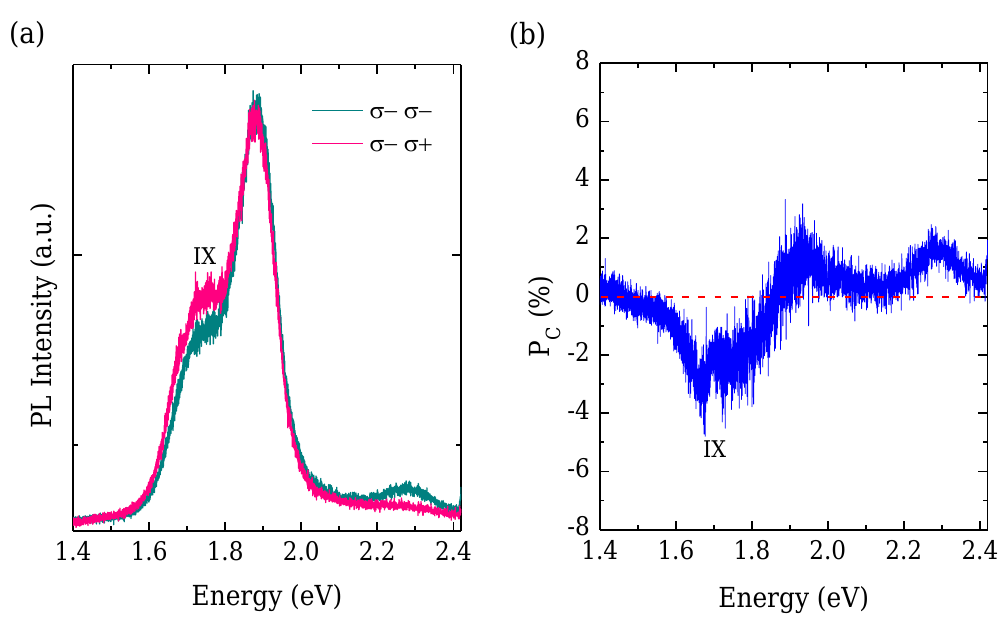}
\caption{Helicity-resolved PL of bilayer WSSe using $\sigma-$ excitation at 2.54 eV with $\sigma-$ and $\sigma+$ detection.}
\end{figure*}
\newpage
\subsection{Helicity-resolved PL measurements on two additional samples}
\begin{figure*}[h!]
 \centering
\includegraphics[width=0.55\linewidth]{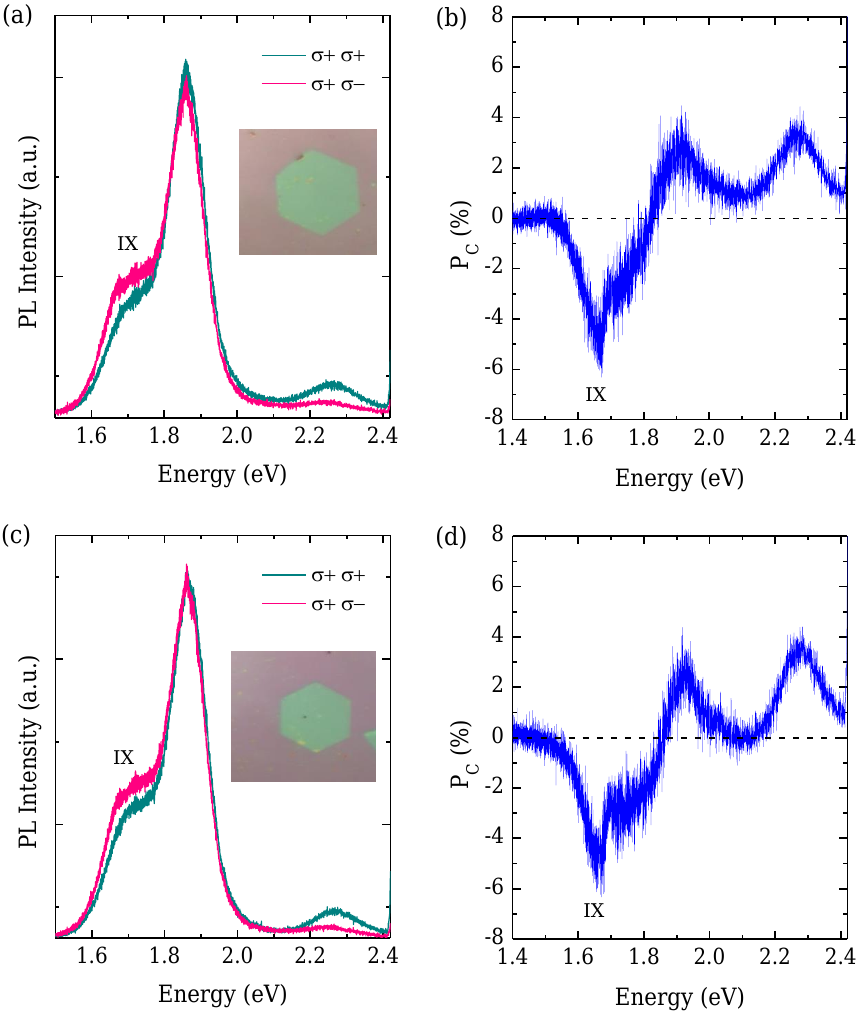}
\caption{ (a), (c) Helicity-resolved PL of two additional samples using $\sigma+$ excitation at 2.54 eV with $\sigma+$ and $\sigma-$ detection. Insets display the optical images of the samples.
(b) and (d) The degree of circular polarization (\%) estimated from a and b.}
\end{figure*}
We performed helicity-resolved PL measurements on two additional samples using $\sigma+$ excitation from the off-resonance 2.54 eV laser at RT, as shown in Fig. S11. The results indicate that the emission helicity of the IX is opposite to that of the intralayer emission, similar to the observation in Fig. 3(a).
\newpage
\subsection{Temperature dependent PL study}
We investigated temperature dependent PL study of bilayer WSSe from temperature range of 300 K to 90 K, as depicted in Fig. S12(a). We selected 2.54 eV excitation to investigate the modulation of emissions for all excitons (A, B and IX) with temperature. Figure 4(c) displays the evolution of the peak energies with temperature which are fitted using the Varshni equation \cite{VARSHNI1967149}:
\begin{equation}
    E_g(T) = E_g(0) - \frac{\alpha T^2}{\beta + T}
\end{equation}
Here, $T$ represents the temperature, $E_g(0)$ denotes the excitonic band gap at $T = 0$ K, and $\alpha$ and $\beta$ are fitting parameters. Based on the fitting results, the following values are obtained: For the IX transition, $E_g(0) = 1.67$ eV, $\alpha = -0.15$ meV/K, and $\beta = 106$ K. For the A exciton, $E_g(0) = 1.91$ eV, $\alpha = 0.2$ meV/K, and $\beta = 237$ K.
\begin{figure*}[h]
 \centering
\includegraphics[width=0.8\linewidth]{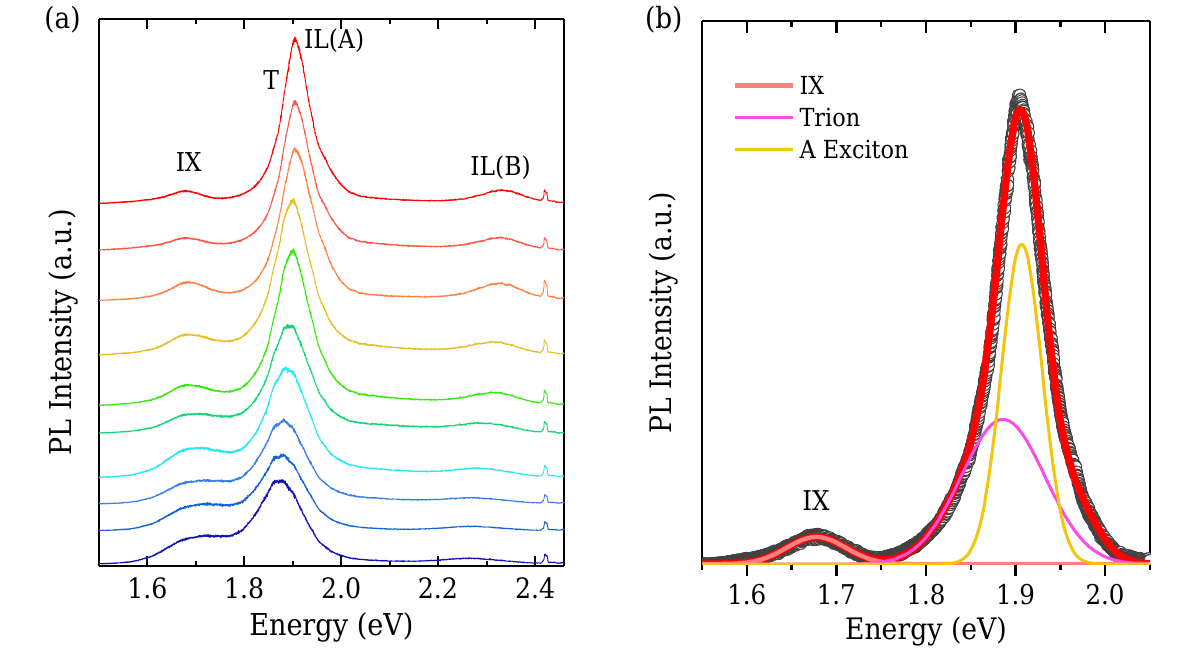}
\caption{(a) Temperature-dependent PL study of bilayer WSSe across a temperature range 300 K to 90 K (bottom to top). (b) Decomposition of different transitions in the PL spectrum at 90 K.}
\end{figure*}
\subsection{Helicity-resolved PL measurements using 1.96 eV excitation ($\sigma$+)}
We performed helicity-resolved PL study using near on-resonance excitation (with A exciton) of 1.96 eV laser with $\sigma +$ helicity from the temperature range 300 K to 90 K. Here, the experiment was conducted by keeping integration time of 2 sec. All the results are displayed in Fig. S13(a) and (b). For the A exciton, we achieve the degree of circular polarization value ($P_C (\%)$) of 40\%, which increases to 72\% at 90 K temperature (Fig. S13(c)).  
\begin{figure*}[h!]
 \centering
\includegraphics[width=1\linewidth]{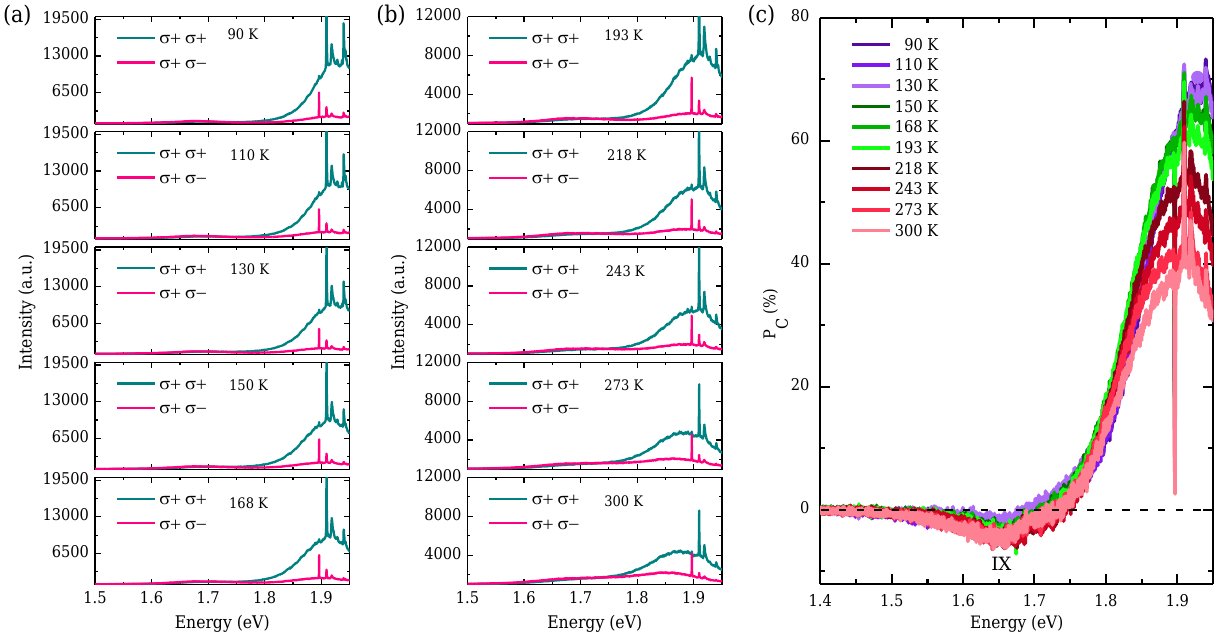}
\caption{Helicity-resolved PL of bilayer WSSe using $\sigma+$ excitation at 1.96 eV (power density: 0.17 mW/$\mu$m$^2$) with $\sigma+$ and $\sigma-$ detection, (a),(b) at various temperatures ranging from 300 K to 90 K, and (c) the corresponding temperature-dependent circular polarization, $P_C (\%)$.}
\end{figure*}
Furthermore, to investigate the modulation of the P$_C$ for both IX and intralayer excitons, we explored power dependent helicity-resolved PL measurements. In this case also, we performed all the experiments by maintaining integration time as 2 sec. Figures S14(a) and (b) display the power-dependent helicity-resolved PL spectra at RT and 90 K, respectively. The extracted values of $P_C (\%)$ are shown in Figs. S14(c) and (d). For the A exciton, we achieve the maximum P$_C$ value of 77\% at 90 K temperature (Fig. S14(d)). 
\begin{figure*}[h!]
 \centering
\includegraphics[width=0.8\linewidth]{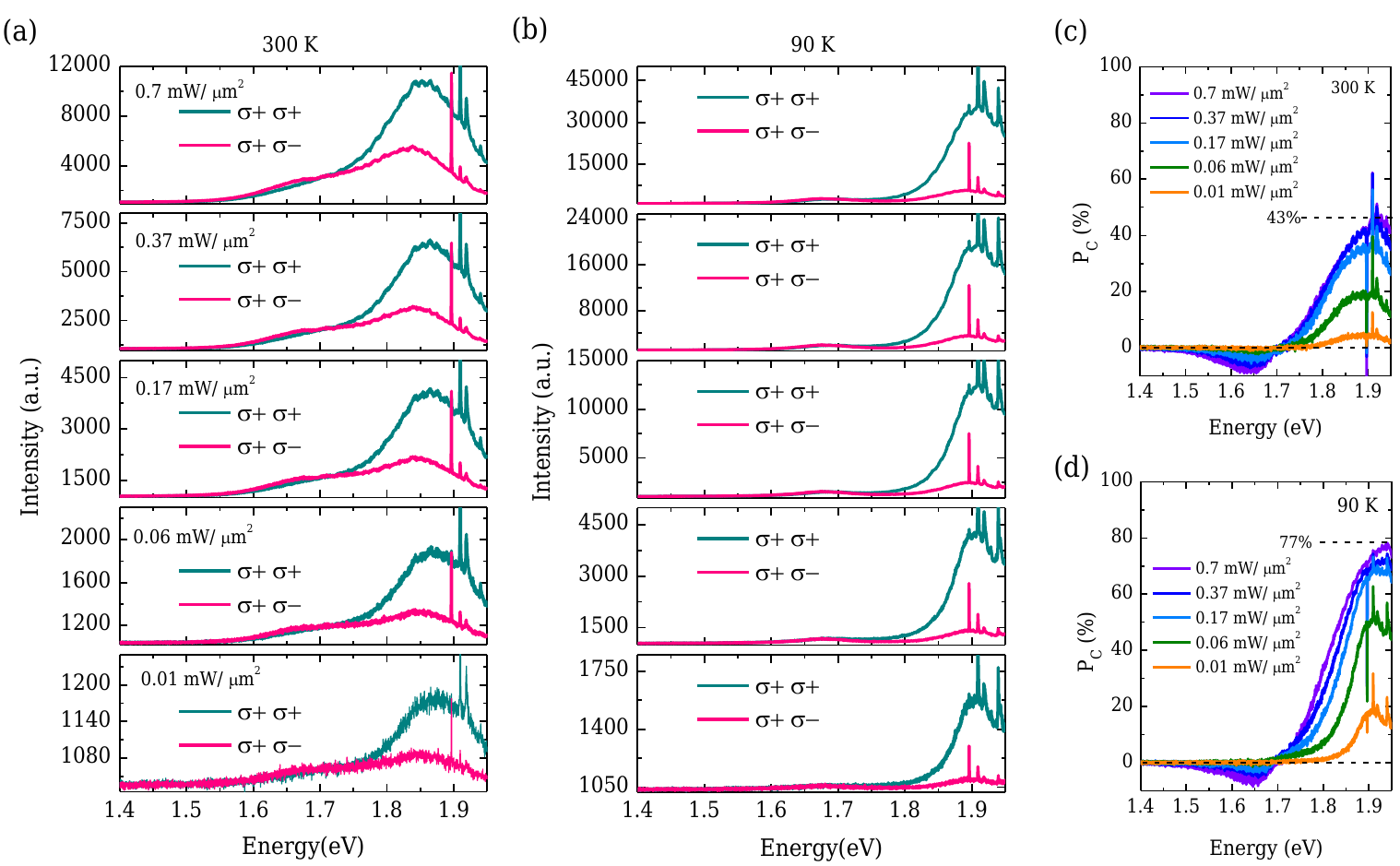}
\caption{Power dependence of helicity-resolved PL of bilayer WSSe using $\sigma+$ excitation at 1.96 eV at (a) RT and (b) 90 K. Extracted degree of circular polarization at (c) RT and (d) 90 K, respectively.}
\end{figure*}
\newpage
\subsection{Excitation energy dependence of circular polarization}
We conducted helicity-resolved PL measurements using 2.54 eV and 1.96 eV lasers to further explore the effect of excitation energy on circularly polarized emission. The measurements were carried out at both RT and 100 K, while keeping the power density (0.17 mW/$\mu$m$^2$) and integration time (10 sec) constant across all the measurements. Figure S15(a), (b) and (d), (e) present the co- and cross-circularly polarized PL spectra at RT and 100 K temperature for 2.54 eV and 1.96 eV excitations, respectively. A negative $P_C$ for IX is observed (Fig. S15(c), (f), top) in both temperatures under off-resonance excitation, consistent with the results shown in Fig. 3(c). Importantly, under resonant conditions (see bottom of Figs. S15(c), (f)), an enhancement of $P_C$ for the IX was observed, with an increase of 2.4 times at RT and 3.2 times at 100 K compared to the 2.54 eV excitation.

\begin{figure*}[h!]
 \centering
\includegraphics[width=0.85\linewidth]{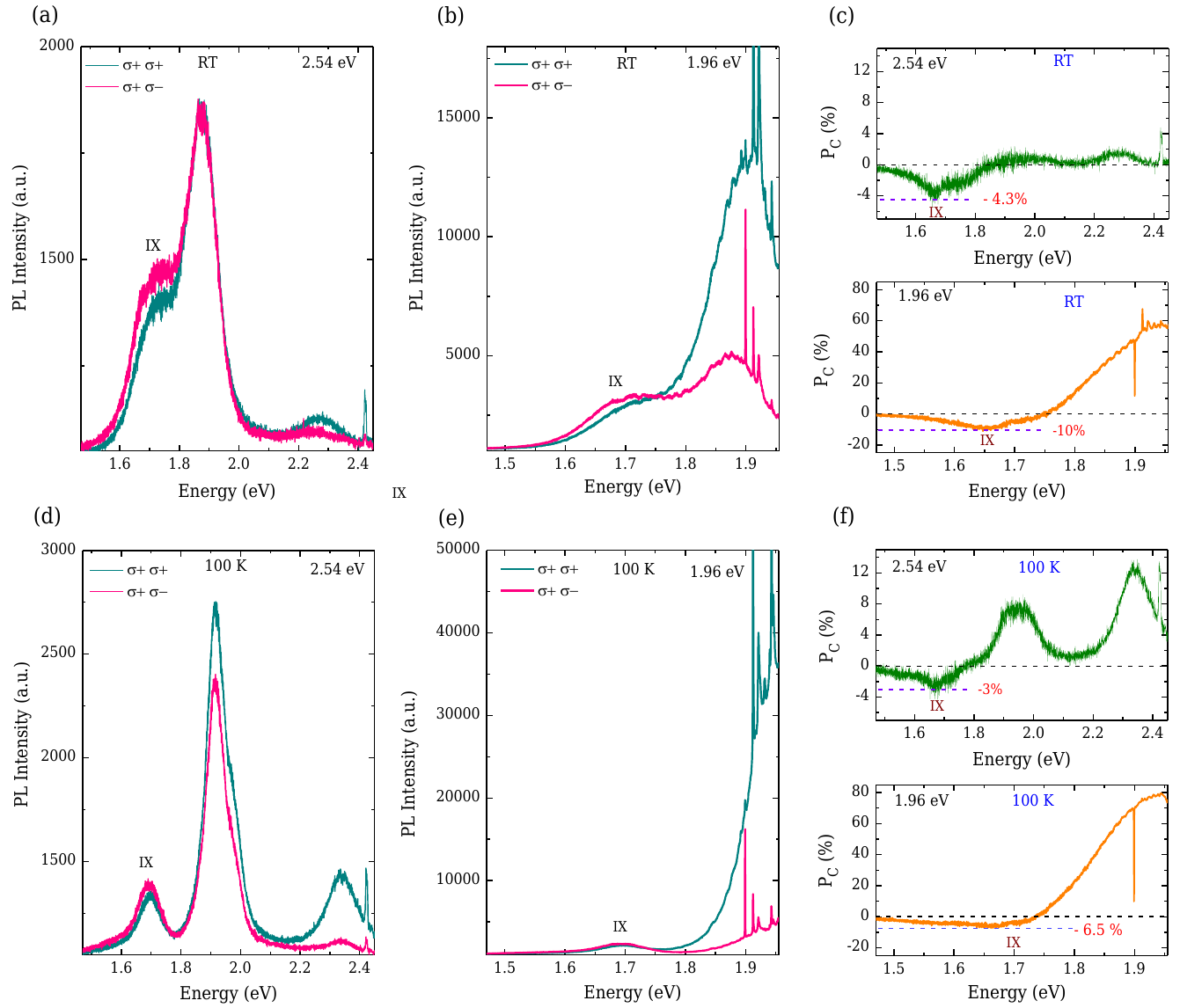}
\caption{Excitation energy dependent circular polarization. Helicity-resolved PL spectra at RT, using (a) 2.54 eV and (b) 1.96 eV excitations. (c) Extracted P$_C$ at RT from (a) and (b). Helicity-resolved PL spectra at 100 K using (d) 2.54 eV and (e) 1.96 eV excitations, and (f) extracted P$_C$ at 100 K from (d) and (e).}
\end{figure*}
\subsection{Model for bilayer}
In our bilayer alloy sample, the electric field is internally generated. The Hamiltonian ($\hat{H}$) for the 2H-stacked TMDC alloy bilayer, in the presence of an out-of-plane electric field E, has the matrix representation \cite{Jones2014}:\\
\begin{equation}
\hat{H}=
\begin{pmatrix}
\Delta-\tau_z s_z\lambda_c+Ed/2 & 0 & at(\tau_z k_x + ik_y) & 0 \\
0 & \Delta+\tau_zs_z\lambda_c - Ed/2 & 0 & at(\tau_z k_x - ik_y) \\
at(\tau_z k_x - ik_y) & 0 & -\tau_z s_z \lambda_v + Ed/2 & t_{\perp} \\
0 & at(\tau_z k_x + ik_y) & t_{\perp} & \tau_z s_z \lambda_v - Ed/2 \\
\end{pmatrix}
\end{equation}
The Hamiltonian, defined close to the Brillouin zone K points, represents a minimal band model in the spirit of the k.p model of TMDC monolayers, with an added interlayer hopping term $t_{\perp}$ for the VB hole. The different symbols are: $a$, the lattice constant, $t$, the intralayer hopping integral, $\Delta$, the monolayer band gap, 2$\lambda_c$ (2$\lambda_v$), the magnitude of the CB (VB) spin-splitting energy, the valley index $\tau_z = $+1 (-1) for the K (-K) valley, $s_z$, the Pauli spin matrix, and $d$, the interlayer distance, respectively. The basis states of the $\hat{H}$ matrix are: $\ket{d^u_{z^2}}$, $\ket{d^l_{z^2}}$, $\frac{1}{\sqrt{2}} (\ket{d^u_{x^2-y^2}} - i\tau_z\ket{d^u_{xy}})$, $\frac{1}{\sqrt{2}} (\ket{d^l_{x^2-y^2}} + i\tau_z\ket{d^l_{xy}})$,  where the superscript $u$ and $l$ denote the upper and lower layer, respectively. The Hamiltonian contains the spin-valley-layer coupling interaction term $- \lambda \sigma_z \tau_z s_z$, where the layer index $\sigma_z = +1 (-1)$ for the upper (lower) layer and $\lambda = \lambda_c (\lambda_v)$ for the CB (VB). At the $\pm$K points, $k_x$, $k_y$ are both zero and one can diagonalize the Hamiltonian to obtain the eigen values and eigenvectors associated with the CB and VB. The CB states are already diagonalized and given by: 
\par
K valley, lower layer, eigenvalues = $ \Delta \pm \lambda_c - Ed/2 $ with respective eigenvectors 
\begin{equation}
\ket{\psi_{CB1, l}}_K = \begin{pmatrix}
0 \\ 1 \\ 0 \\ 0 \\
\end{pmatrix} \otimes \ket{\uparrow},
\ket{\psi_{CB2, l}}_K = \begin{pmatrix}
0 \\ 1 \\ 0 \\ 0 \\
\end{pmatrix} \otimes \ket{\downarrow}.
\end{equation}
\par
K valley, upper layer, eigenvalues = $ \Delta \pm \lambda_c + Ed/2 $ with respective eigenvectors
\begin{equation}
\ket{\psi_{CB1, u}}_K = \begin{pmatrix}
1 \\ 0 \\ 0 \\ 0 \\
\end{pmatrix} \otimes \ket{\downarrow},
\ket{\psi_{CB2, u}}_K = \begin{pmatrix}
1 \\ 0 \\ 0 \\ 0 \\
\end{pmatrix} \otimes \ket{\uparrow}.
\end{equation}
\par
-K valley, lower layer, eigenvalues = $ \Delta \pm \lambda_c - Ed/2 $ with respective eigenvectors
\begin{equation}
\ket{\psi_{CB1, l}}_{-K} = \begin{pmatrix}
0 \\ 1 \\ 0 \\ 0 \\
\end{pmatrix} \otimes \ket{\downarrow},
\ket{\psi_{CB2, l}}_{-K} = \begin{pmatrix}
0 \\ 1 \\ 0 \\ 0 \\
\end{pmatrix} \otimes \ket{\uparrow}.
\end{equation}
\par
-K valley, upper layer, eigenvalues = $ \Delta \pm \lambda_c + Ed/2 $ with respective eigenvectors
\begin{equation}
\ket{\psi_{CB1, u}}_{-K} = \begin{pmatrix}
1 \\ 0 \\ 0 \\ 0 \\
\end{pmatrix} \otimes \ket{\uparrow},
\ket{\psi_{CB2, u}}_{-K} = \begin{pmatrix}
1 \\ 0 \\ 0 \\ 0 \\
\end{pmatrix} \otimes \ket{\downarrow}.
\end{equation}
\begin{figure*}[t!]
 \centering
\includegraphics[width=0.5\linewidth]{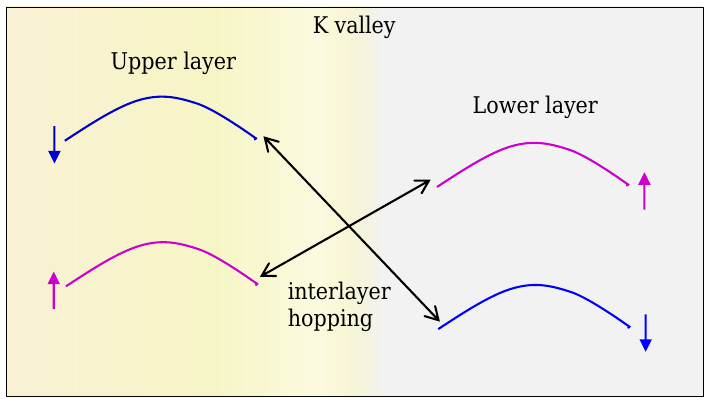}
\caption{Schematic energy diagram of the valence band for bilayer WSSe.}
\end{figure*}
The VB eigenvalues and eigenstates are obtained by diagonalizing the 2$\times$2 matrix at the bottom corner of the Hamiltonian matrix (Eq. S2). In Fig. S16, a schematic diagram of the VB structure in the presence of an out-of-plane electric field is shown with hybridization between the lower and upper layer bands. Due to the large magnitude (a few hundred meV) of 2$\lambda_v$, the VB energy splitting, we focus attention on the higher energy VB eigenstate. The CB (VB) states under consideration are the electron (hole) states generated via laser excitation. The eigenvectors of the Hamiltonian matrix representing the hole states, largely localized in the lower layers of the K and -K valleys, are:
\begin{equation}
\ket{K, \uparrow} = \begin{pmatrix}
0 \\ 0 \\ \alpha_2(E) \\ \sqrt{1 - |\alpha_2(E)|^2} \\
\end{pmatrix} \otimes \ket{\uparrow},
\end{equation}
\begin{equation}
\ket{-K, \downarrow} = \begin{pmatrix}
0 \\ 0 \\ \alpha_2(E) \\ \sqrt{1 - |\alpha_2(E)|^2} \\
\end{pmatrix} \otimes \ket{\downarrow}.
\end{equation}
Here, the coefficient 
\begin{equation}
    \alpha_2(E) = \frac{t_{\perp}}{\sqrt{(E_l + \lambda _v - Ed/2)^2 + |t_{\perp}|^2}}.
\end{equation}
The eigenvalue is 
\begin{equation}
    E_l = \sqrt{(\lambda_v - Ed/2)^2 + |t_{\perp}|^2}.
\end{equation}
\par
Similarly, the hole states predominantly localized in the upper layer are represented by the eigenvectors
\begin{equation}
\ket{K, \downarrow} = \begin{pmatrix}
0 \\ 0 \\ \sqrt{1 - |\alpha_1(E)|^2} \\ \alpha_1(E) \\
\end{pmatrix} \otimes \ket{\downarrow},
\end{equation}
\begin{equation}
\ket{-K, \uparrow} = \begin{pmatrix}
0 \\ 0 \\ \sqrt{1 - |\alpha_1(E)|^2} \\ \alpha_1(E) \\
\end{pmatrix} \otimes \ket{\uparrow}.
\end{equation}
Here, the coefficient   
\begin{equation}
    \alpha_1(E) = \frac{t_{\perp}}{\sqrt{(E_u + \lambda _v + Ed/2)^2 + |t_{\perp}|^2}}.
\end{equation}
The eigenvalue is 
\begin{equation}
    E_u = \sqrt{(\lambda_v + Ed/2)^2 + |t_{\perp}|^2}
\end{equation}
The expressions for $\alpha_1(E)$, $\alpha_2(E)$, from Eqs. (S9) and (S13), can be reexpressed as
\begin{equation}
|\alpha_1(E)|^2 = \frac{1}{2} - \frac{\lambda_v + Ed/2}{2E_u}
\end{equation}
\begin{equation}
|\alpha_2(E)|^2 = \frac{1}{2} - \frac{\lambda_v - Ed/2}{2E_l}.
\end{equation}
\par
The estimates of $2\lambda_c$, from band structure calculations, are 27 meV (WS$_2$) and 38 meV (WSe$_2$) respectively \cite{PhysRevB.88.245436}. For the bilayer alloy material, WS$_{2x}$Se$_{2(1-x)}$,($x$ = 0.88), we apply Anderson’s rule \cite{5392495, PhysRevLett.55.418, doi:10.1021/acs.nanolett.5b03662} to obtain the value,
\begin{equation}
2\lambda_c (WS_{2x}Se_{2(1-x)}) = x.2\lambda_c (WS_{2}) + (1-x).2\lambda_c (WSe_{2(1-x)})
\end{equation}
which is calculated as $2\lambda_c$ = 0.02832 eV. The  $2\lambda_v$ values from band structure calculations are 421 meV (WS$_2$) and 456 meV (WSe$_2$) respectively \cite{Gong2013, Jones2014}. Using Anderson’s rule, the estimate for the bilayer alloy is $2\lambda_v$ = 0.4252 eV. In the case of the A exciton in the $–K$ valley of Fig. 3(d), the energy of the CB1 band in the upper layer, occupied by the photogenerated electrons, is $\Delta + \lambda_c + Ed/2$. In the case of the IX, the energy of the CB2 band in the lower layer is $\Delta - \lambda_c - Ed/2$. The holes in both the cases are in the same VB state $\ket{-K, \uparrow}$. The difference in the peak position of the A exciton and IX is thus $2\lambda_c + Ed$.

\subsection{Fine structure of interlayer exciton}
In the case of monolayer TMDCs, a bright exciton state involves a VB and CB state in which
spin is conserved whereas opposite-spin states constitute a dark exciton. In a theoretical study,
Yu et al. \cite{Yu_2018} have shown that in the case of heterobilayers/homobilayers, more general optical
selection rules, depending on the local atomic registry, are applicable allowing for the formation
of bright IX exciton states via spin-flipping transitions. For example, in Fig. S17, two interlayer
excitons, labeled IX1 (spin-conserved) and IX2 (spin-flipped), could be optically active. The fine
structure of the IX has recently been detected in the PL emission signals of heterostructures at
low temperatures \cite{Yu:20, Durmuş2024, doi:10.1021/acs.nanolett.9b04528}. Though we did not directly observe the two-peaked structure of the
IX in the PL spectrum, a data analysis of the PL intensity at 90 K shows that a double-peaked
structure gives a good fit to the asymmetric shape of the experimental data rather than a single
peak (Figs. S18(b) and S18(c)). The difference in the peak energies is $\sim$26 meV close to the CB
spin-splitting energy of $\sim$28 meV for the bilayer alloy material. This feature, uncovered through
a data analysis, is consistent with recent experimental observations \cite{Yu_2018, Yu:20, Durmuş2024}.

\begin{figure*}[h!]
 \centering
\includegraphics[width=0.8\linewidth]{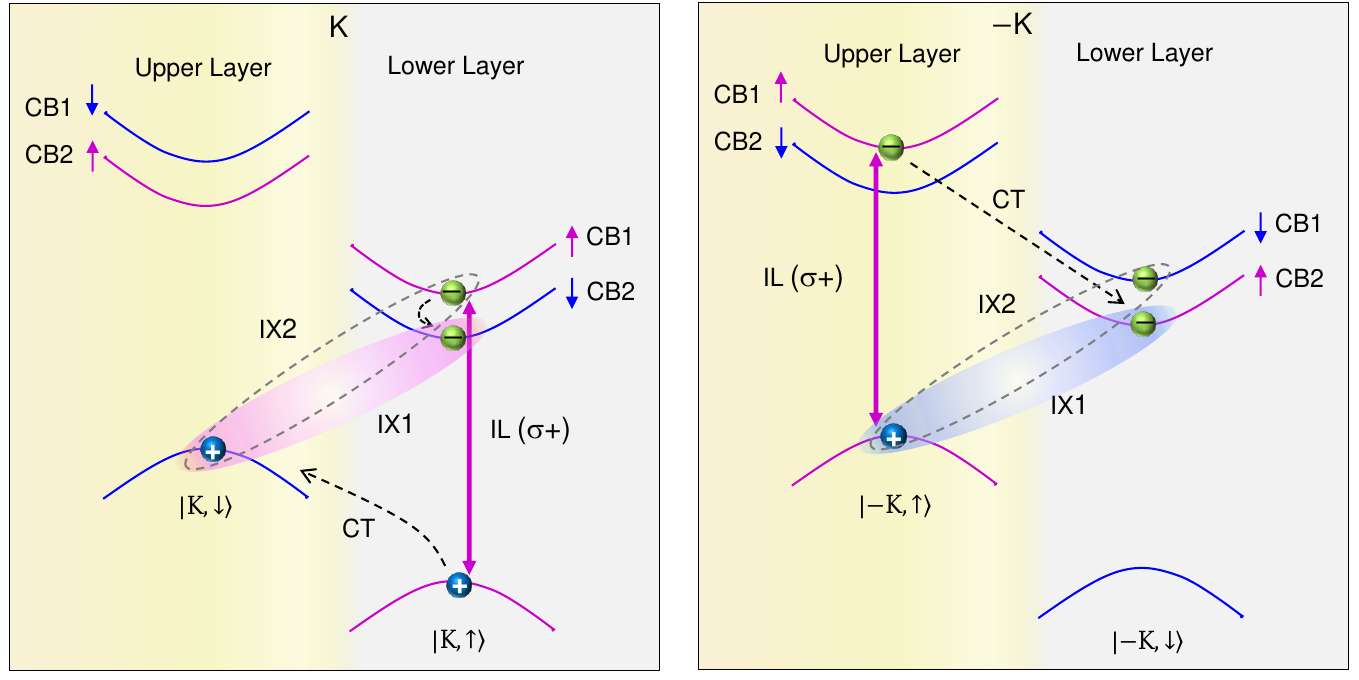}
\caption{Schematic energy diagram of
bilayer WSSe considering two interlayer transitions.}
\end{figure*}
\begin{figure*}[h!]
 \centering
\includegraphics[width=1\linewidth]{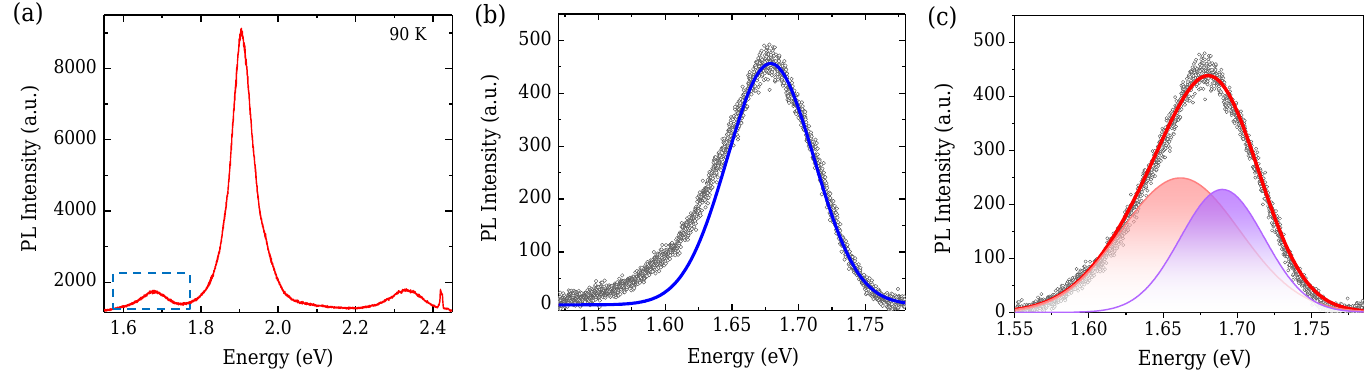}
\caption{(a) PL spectrum of bilayer WSSe taken at 90 K. (b) Deconvolution of the IX emission region (blue dotted box in a) using single peak fitting, and (c) two peaks fitting.}
\end{figure*}

\newpage
\def\bibsection{\subsection*{\refname}} 
%

\subsection*{Acknowledgements} 
 S.M. thanks the Council of Scientific \& Industrial Research (CSIR), India, for financial support through the NET-SRF award (File No. 09/015(0531)/2018-EMR-I). C.N. acknowledges the INSPIRE Fellowship Programme, DST, Government of India, for her research fellowship (Reg. No. IF180057). I.B. expresses gratitude for the support from NASI, Allahabad, India, under the Honorary Scientist Scheme. A.S. is grateful for financial assistance from the Science and Engineering Research Board (SERB), India (File No. EMR/2017/002107). We sincerely acknowledge Shubhadip Moulick (S. N. Bose National Centre for Basic Sciences) for his assistance in device fabrication. 
 \subsection*{Author contributions}
 S.M. and A.S. conceived and designed the experiments. S.M. and C.N. synthesized the 2D alloy TMDC and performed the experiments. S.M., C.N., and A.S. analyzed the experimental data. P.C. and A.N.P. fabricated the device for gate voltage-dependent study. I.B. performed the theoretical model calculations and contributed towards conceptualization and physical interpretations. S.M., C.N., I.B., and A.S. contributed to the writing of the manuscript.
\end{document}